\documentclass[
amsmath,
prd,nofootinbib,floatfix,12pt,
]{revtex4}

\usepackage{graphicx}
\usepackage{setspace}
\usepackage{bm}

\newcommand\beq{\begin{eqnarray}}
\newcommand\eeq{\end{eqnarray}}
\def\lsim{\mathrel{\rlap{\lower4pt\hbox{$\sim$}}
    \raise1pt\hbox{$<$}}}                
\def\gsim{\mathrel{\rlap{\lower4pt\hbox{$\sim$}}
    \raise1pt\hbox{$>$}}}            

\allowdisplaybreaks
\newcommand\Fbar{\overline{F}}
\newcommand\Ibar{\overline{I}}
\newcommand\Abar{\overline{A}}

\newcommand\psibar{\overline{\psi}}

\newcommand\lnbar{\overline{\ln}}
\newcommand\zetathree{{\zeta_3}}
\newcommand\zetatwo{{\zeta_2}}

\newcommand\MSbar{$\overline{\rm{MS}}$ }
\newcommand\DRbarprime{$\overline{\rm{DR}}'$ }
\newcommand\cc{\mbox{c.c.}}

\def\gptwo{g^{\prime 2}}   
\def\gpfour{g^{\prime 4}}

\begin{document}

\renewcommand{\theequation}{\arabic{section}.\arabic{equation}}
\renewcommand{\thefigure}{\arabic{section}.\arabic{figure}}
\renewcommand{\thetable}{\arabic{section}.\arabic{table}}

\title{\large Effective potential at three loops}

\author{Stephen P.~Martin}
\affiliation{Department of Physics, Northern Illinois University, DeKalb IL 60115}

\begin{abstract}\normalsize \baselineskip=15pt 
I present the effective potential at three-loop order for a general 
renormalizable theory, using the \MSbar renormalization scheme and Landau gauge fixing. 
As applications and illustrative points of reference, the 
results are specialized to the supersymmetric Wess-Zumino model
and to the Standard Model. In each 
case, renormalization group scale invariance provides a consistency check. 
In the Wess-Zumino model, the required
vanishing of the minimum vacuum energy yields an additional check. 
For the Standard Model, I carry out the resummation of Goldstone boson contributions, which provides yet more opportunities for non-trivial checks,
and obtain the minimization condition for the Higgs vacuum expectation value
at full three-loop order. 
An infrared divergence due to doubled photon propagators appears in the 
three-loop Standard Model effective potential,
but it does not affect the minimization condition or physical observables
and can be eliminated by resummation.
\end{abstract}

\maketitle

\vspace{-0.4in}
\baselineskip=14.45pt 

\tableofcontents

\baselineskip=16pt
\newpage

\section{Introduction \label{sec:intro}}
\setcounter{equation}{0}
\setcounter{figure}{0}
\setcounter{table}{0}
\setcounter{footnote}{1}

The effective potential \cite{Coleman:1973jx,Jackiw:1974cv,Sher:1988mj} is a useful 
tool for understanding spontaneous symmetry breaking in quantum field theories. 
It can be defined in perturbation theory, and calculated, by expanding the scalar fields
appearing in the Lagrangian about constant background values $\phi$, and then
summing the one-particle-irreducible vacuum (no external legs) Feynman diagrams,
using propagator masses and interaction vertices that depend on the background scalar fields.
The full two-loop effective potential has been obtained for the Standard Model 
in the \MSbar scheme and Landau gauge
by Ford, Jack, and Jones in ref.~\cite{Ford:1992pn}, and in general theories
(including softly broken supersymmetric ones, which use a different regulator based on
dimensional reduction) 
in ref.~\cite{Martin:2001vx}. The three-loop effective
potential for the Standard Model has been found in the approximation
that the the QCD and top Yukawa couplings are larger than all other couplings,
in ref.~\cite{Martin:2013gka}, and the four-loop contribution only at leading order in
QCD \cite{Martin:2015eia}.

One application of the effective potential is to study the 
stability properties of our vacuum state in the Standard Model \cite{Sher:1988mj},
\cite{Lindner:1988ww}-\cite{Espinosa:2016nld}
and extensions of it. This has attracted great interest recently due to the apparent 
proximity of the Higgs boson self-coupling to the critical value associated with metastability.
 
Another important use of the effective potential 
is to relate the vacuum expectation value (VEV) of 
the symmetry breaking scalar field(s) to the Lagrangian parameters, including 
the negative Higgs squared mass parameter. 
Note that the VEV can be defined as the value of
the constant scalar background fields that minimizes either the tree-level potential
or the full effective potential. Choosing the first definition, with the VEV
as the minimum of the tree-level potential, has the advantages of providing
gauge-invariant running masses, and allowing for checks in subsequent calculations
of other quantities by varying
the gauge-fixing parameter. However, it requires the inclusion of tadpole diagrams
in those calculations (see for example \cite{Fleischer:1980ub}-\cite{Bednyakov:2016onn}), 
which causes inverse powers of the Higgs coupling 
to appear in perturbation theory. 

By defining the VEV as the minimum of the full 
effective potential, the sum of all tadpole graphs automatically vanishes, 
and inverse powers of the Higgs self-coupling do not occur, 
so that in calculations of other quantities, perturbation
theory converges faster.
The price to be paid for using this ``tadpole-free" scheme 
is that the resulting VEV is dependent on the gauge-fixing choice, and therefore 
so are the running \MSbar
masses of the particles. This is of course not a real problem,
because the VEV and the running masses are not physical observables. 
Calculations in this approach are simplest in Landau gauge, where there is no mixing
between Goldstone bosons and vector gauge bosons and the gauge-fixing parameter is not
renormalized. (Ref.~\cite{Degrassi:2014sxa} is a good example of using 
a tadpole-free scheme but in Feynman gauge.)
Although fixing to Landau gauge 
precludes obtaining checks from varying the gauge-fixing parameter, there are checks of
similar power from cancellations in observable quantities between Goldstone bosons 
and the unphysical components of vector degrees of freedom.
The tadpole-free pure \MSbar scheme has been used to 
calculate the complex pole masses of the 
Higgs \cite{Martin:2014cxa}, 
$W$ \cite{Martin:2015lxa}, 
and $Z$ \cite{Martin:2015rea} bosons, and the top quark \cite{Martin:2016xsp},
to full two-loop order, in terms of the \MSbar Lagrangian parameters, 
using notation and computational methods consistent with the present paper.

In this paper, I will obtain the three-loop effective potential for a general renormalizable quantum field theory in four dimensions, using Landau gauge fixing
and the \MSbar scheme 
\cite{Bardeen:1978yd,Braaten:1981dv}
based on dimensional regularization
\cite{Bollini:1972ui,Ashmore:1972uj,Cicuta:1972jf,tHooft:1972fi,tHooft:1973mm}. 
In the following, 
$1/16\pi^2$ is used as a loop expansion parameter, so that the
effective potential is written as:
\beq
V_{\rm eff}(\phi) &=& V^{(0)} 
+ \frac{1}{16 \pi^2} V^{(1)}
+ \frac{1}{(16 \pi^2)^2} V^{(2)}
+ \frac{1}{(16 \pi^2)^3} V^{(3)}
+ \ldots
\eeq
As is well-known, 
the contribution $V^{(\ell)}$ is obtained as the sum of one-particle-irreducible
$\ell$-loop vacuum Feynman diagrams, using propagator masses and vertices that depend on
the constant background scalar field(s) $\phi$. First derivatives of the 
effective potential correspond to tadpole diagrams involving the scalar fields,
and so working at the minimum of $V(\phi)$ guarantees that the sum of tree-level
and loop-corrected tadpoles vanishes,
and therefore tadpoles need not be included in other calculations. The new results for the
contributions to $V^{(3)}$ will be presented in section \ref{sec:mainresults}.
As illustrative applications of the general results, I will specialize them to the
cases of the supersymmetric Wess-Zumino model and the Standard Model, in sections 
\ref{sec:WZ} and \ref{sec:SM} respectively. Many of the results obtained below are too lengthy to show in print, and so are presented instead in ancillary electronic files in 
forms suitable for use with computers.

An important way of checking a calculation of the effective potential is by requiring
renormalization group invariance, provided that the pertinent beta functions have
already been calculated to the corresponding order by other means. The requirement
that $V_{\rm eff}$ does not depend on the choice of the \MSbar renormalization
scale $Q$ can be written as
\beq
Q\frac{dV_{\rm eff}}{dQ} &=& \biggl ( Q \frac{\partial}{\partial Q} 
+ \sum_{X} \beta_X \frac{\partial}{\partial X} \biggr )
V_{\rm eff} = 0 .
\label{eq:QdVdQgen}
\eeq
where $X$ runs over all of the independent Lagrangian parameters, 
including the background scalar field(s).
The beta function for a background scalar field 
$\phi$ is related to its anomalous dimension $\gamma$ 
by $\beta_\phi = -\phi \gamma$.
The loop expansions for the beta functions of $X$ can be written:
\beq
\beta_X &=&
\frac{1}{16 \pi^2} \beta_X^{(1)}
+\frac{1}{(16 \pi^2)^2} \beta_X^{(2)}
+\frac{1}{(16 \pi^2)^3} \beta_X^{(3)} + \ldots.
\label{eq:betaX}
\eeq
Then it follows that at each loop order $\ell = 1,2,3,\ldots$, one must have:
\beq
Q \frac{\partial}{\partial Q} V^{(\ell)} + 
\sum_{n=0}^{\ell-1} 
\biggl (\sum_{X} \beta_{X}^{(\ell - n)} 
\frac{\partial}{\partial X} V^{(n)} \biggr)
&=& 0 .
\label{eq:QdQVexpanded}
\eeq
This will be applied below as a check in both the Wess-Zumino model
and Standard Model.

\section{Conventions and setup \label{sec:setup}}
\setcounter{equation}{0}
\setcounter{figure}{0}
\setcounter{table}{0}
\setcounter{footnote}{1}

The conventions and notations for this paper related to the effective potential
and to two-component fermions 
generally follow refs.~\cite{Martin:2001vx} and
\cite{Dreiner:2008tw} respectively, 
with some minor cosmetic variations. After expansion about the constant 
background scalar field(s),
the Lagrangian can be written without loss of generality
in terms of real scalars $R_j$, 
real vectors $A^{\mu a}$, and left-handed two-component fermion fields $\psi_I$, with 
background-field-dependent masses and interaction couplings.
(In many cases, complex bosonic fields with well-defined charges 
could be used, but in order to present results in a general way, I take advantage of
the fact that they can always be decomposed into real and imaginary parts.) 
For fermion fields that carry conserved charges, 
it is most convenient to use pairs of 2-component
left-handed fields $\psi_I$ and $\psi_{I'}$ with opposite charges and therefore
a purely off-diagonal Dirac mass $M^{II'}$, so that the
common squared mass for both fields is $M_I^2 = M_{I'}^2 \equiv |M^{II'}|^2$. 
This means that the two-component fermion fields are always 
squared mass eigenstates but sometimes not mass eigenstates.

The squared-mass eigenstate fields 
are therefore labeled by indices $j,k,l,m,n,p$ for real scalars,
$a,b,c,d,e,f$ for real vectors, and $I,J,K,L$ for two-component fermions, with the
understanding that $I', J', K', L'$ are used to denote the corresponding mass partners
when they form a Dirac pair, and with $I'=I$ for a fermion with a Majorana-type mass.
As a convention, repeated indices are always taken to be summed over. 

The most general
interaction Lagrangian for a renormalizable theory can be written in terms of 
background-field-dependent couplings as (using a metric of signature $-$,$+$,$+$,$+$): 
\beq
{\cal L} &=& 
-\frac{1}{6} \lambda^{jkl} R_j R_k R_l
-\frac{1}{24} \lambda^{jklm} R_j R_k R_l R_m
-\frac{1}{2} \left ( Y^{jIJ} R_j \psi_I \psi_J + {\rm c.c.} \right )
\nonumber \\ && 
+ g^{aJ}_I A^{\mu a} \psi^{\dagger I} \overline{\sigma}_\mu \psi_J 
- g^{ajk} A^{\mu a}R_j \partial_\mu R_k
- \frac{1}{4} g^{abjk} A_\mu^a A^{\mu b} R_j R_k
- \frac{1}{2} g^{abj} A_\mu^a A^{\mu b} R_j
\nonumber \\ &&
- g^{abc} A^{\mu a} A^{\nu b} \partial_\mu A_\nu^c
- \frac{1}{4} g^{abe} g^{cde} A^{\mu a} A^{\nu b} A^c_\mu A^d_\nu
- g^{abc} A^{\mu a} \omega^b \partial_\mu \overline{\omega}^c
.
\label{eq:Linteractions}
\eeq
Here $\lambda^{jkl}$ and $\lambda^{jklm}$ are real scalar interactions that 
are totally symmetric under interchange of all indices,
$Y^{jIJ}$ are Yukawa couplings that are symmetric under interchange of $I,J$, 
and $g^{aJ}_I$ are vector interactions with fermions, and
$g^{ajk}$, $g^{abjk}$, and $g^{abj}$ are vector interactions with scalars, and
$g^{abc}$ are vector self-interactions. 
Note that the sign of $g^{abc}$ has been flipped compared to 
the notation of ref.~\cite{Martin:2001vx}; this has no impact on the results of
that reference, because at two-loop order $g^{abc}$ only appears in pairs.
Because the scalars and vectors are real,
the heights of their indices have no significance, and are chosen for typographical
convenience. As a convention, flipping the heights of all fermion indices corresponds 
to complex conjugation, so that
\beq
Y_{jIJ} &\equiv& \left (Y^{jIJ} \right )^*,
\\
M_{II'} &\equiv& \bigl (M^{II'}\bigr )^* ,
\eeq
and
\beq
g^{aI}_J &=& \left (g^{aJ}_I \right )^* .
\eeq
All of the couplings with names involving $g$ have their origin as gauge couplings. 
In Landau gauge, the ghost fields $\omega^a$ and $\overline \omega^a$ are massless.
Note that the vector cubic, vector quartic, and vector-ghost-antighost interactions
are all written in terms of a common, totally anti-symmetric, $\phi$-dependent, coupling $g^{abc}$.
The vector-vector-scalar couplings $g^{abj}$ are symmetric 
under interchange of the vector indices
$a,b$, and the vector-scalar-scalar couplings $g^{ajk}$ are 
anti-symmetric under interchange of the scalar indices $j,k$.
Note also that the vector-vector-scalar-scalar couplings are not independent;
they can always be written in terms
of the vector-scalar-scalar couplings, according to 
\beq
g^{abjk} = g^{ajl} g^{bkl} + g^{akl} g^{bjl} .
\label{eq:VVSScouplings}
\eeq
Ref.~\cite{Martin:2001vx} did not mention or exploit this fact, as it 
leads to only a slight simplification at two-loop order, 
but it is used extensively below.

In the following, the names of fields or the corresponding indices
will be used as synonyms for the corresponding
squared mass arguments used in loop integral functions. For example, we can 
write the well-known 1-loop effective potential
in the \MSbar scheme and Landau gauge as simply
\beq
V^{(1)} &=& \sum_j f(j) - 2 \sum_I f(I) + 3 \sum_a f_V(a)
,
\label{eq:V1loopgeneral}
\eeq
where
\beq
f(x) &=& x A(x)/4 - x^2/8 \>=\> \frac{x^2}{4} (\lnbar(x) - 3/2),
\label{eq:deff}
\\
f_V(x) &=& x A(x)/4 + x^2/24 \>=\>  \frac{x^2}{4} (\lnbar(x) - 5/6) 
.
\label{eq:deffV}
\eeq
Here,
\beq
\lnbar(x) \equiv \ln(x/Q^2)
\eeq
where $Q$ is the \MSbar renormalization scale, $x$ is the squared mass argument,
and
\beq
A(x) &=& x \lnbar(x) -x
\label{eq:defA}
\eeq
is a one-loop integral basis function (which was denoted by $J(x)$ in 
ref.~\cite{Martin:2001vx}). The two-loop contribution to the effective potential
is
\beq
V^{(2)} &=& 
\frac{1}{12} (\lambda^{jkl})^2 f_{SSS}(j,k,l) 
  +\frac{1}{8} \lambda^{jjkk} f_{SS}(j,k)
\nonumber \\ &&
  +\frac{1}{2} Y^{jIJ} Y_{jIJ} f_{FFS}(I,J,j)
  +\frac{1}{4} \left (Y^{jIJ} Y^{jI'J'} M_{II'} M_{JJ'} + \cc \right ) 
  f_{\Fbar\Fbar S}(I,J,j)
\nonumber \\ &&
  +\frac{1}{4} (g^{ajk} )^2  f_{VSS}(a,j,k)
  +\frac{1}{4} (g^{abj} )^2  f_{VVS}(a,b,j)
\nonumber \\ &&
  +\frac{1}{2} g^{aJ}_I g^{aI}_J f_{FFV}(I,J,a)  
  +\frac{1}{2} g^{aJ}_I g^{aJ'}_{I'} M^{II'} M_{JJ'} f_{\Fbar\Fbar V}(I,J,a)
\nonumber \\ &&
  + \frac{1}{12} (g^{abc})^2 f_{\mbox{gauge}}(a,b,c)  
,
\label{eq:V2loopgeneral}
\eeq
in terms of two-loop integral functions 
$f_{SSS}$, $f_{SS}$, $f_{FFS}$, $f_{\Fbar\Fbar S}$,
$f_{VSS}$, $f_{VVS}$, $f_{FFV}$, $f_{\Fbar\Fbar V}$, and $f_{\rm gauge}$
that were originally computed (in a different notation) by Ford, Jack,
and Jones in ref.~\cite{Ford:1992pn} in the context of the Standard Model. 
They were given in ref.~\cite{Martin:2001vx} in the 
context of a general renormalizable theory,
in the notation of the present paper, with one exception; in 
eq.~(\ref{eq:V2loopgeneral}) above,
I have used eq.~(\ref{eq:VVSScouplings}) to combine the terms
that involved the functions 
$f_{SSV}$ and $f_{VS}$ in ref.~\cite{Martin:2001vx}, by defining a new function
\beq
f_{VSS}(x,y,z) &\equiv& f_{SSV}(y,z,x) + f_{VS}(x,y) + f_{VS}(x,z) .
\label{eq:deffVSS}
\eeq
Explicitly,
\beq
f_{VSS}(x,y,z) &=& \bigl [  
-\lambda(x,y,z) I(x,y,z) + (y-z)^2 I(0,y,z)  
+ (2x-y+z) A(x) A(y) \phantom{xxxxx}
\nonumber \\ &&
+ (2x+y-z) A(x) A(z)
\bigr ]/x 
+ A(y) A(z) 
+ 2 (y+z-x/3) A(x) 
\nonumber \\ &&
+ 2 x [A(y) + A(z)] 
\label{eq:deffVSSxyz}
\eeq
for $x\not= 0$, and
\beq
f_{VSS}(0,y,z) &=& 3 (y+z) I(0,y,z) + 3 A(y) A(z) -2 y A(y) - 2 z A(z) + (y+z)^2,
\phantom{xxx}
\label{eq:deffVSS0yz}
\eeq
where $I(x,y,z)$ is a two-loop basis integral used in 
refs.~\cite{Ford:1992pn,Martin:2001vx}, and defined in the latter reference 
in the notation appropriate for the present paper, and
\beq
\lambda(x,y,z) \equiv x^2 + y^2 + z^2 - 2 x y - 2 x z - 2 y z
\eeq
is the usual triangle function.

\begin{figure}[t]
\begin{center}
\includegraphics[width=0.99\linewidth,angle=0]{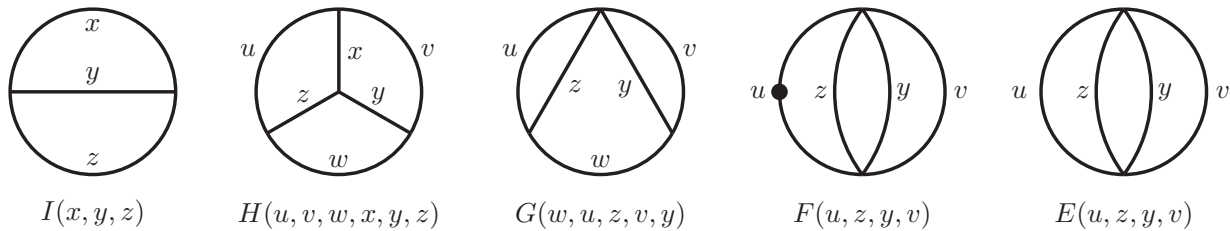}
\end{center}
\vspace{-0.25cm}
\begin{minipage}[]{0.96\linewidth}
\caption{\label{fig:basistopologies}
The topologies for the 2-loop and 3-loop basis vacuum 
integral functions used in this paper. The large dot
in $F(u,z,y,v)$ corresponds to a derivative with respect to the $u$ squared mass argument.
The function $\Fbar(u,z,y,v)$ is the same as $F(u,z,y,v)$ but with a subtraction
to render it infrared 
finite in the limit $u \rightarrow 0$. In each case, counterterms have been included to make the integrals finite as the ultraviolet regulator
$\epsilon \rightarrow 0$.
See ref.~\cite{Martin:2016bgz} for the precise definitions and more information.}
\end{minipage}
\end{figure}

In the following, I will present the three-loop contribution to the effective potential
in terms of three-loop vacuum integral functions using the notation of
ref.~\cite{Martin:2016bgz}, which also provides a computer code {\tt 3VIL} 
for their numerical evaluation\footnote{See also 
refs.~\cite{Freitas:2016zmy,Bauberger:2017nct} for a different approach 
to numerical computation of
the 3-loop vacuum basis integrals, based on dispersion relations. Also, 
refs.~\cite{Davydychev:1992mt}-\cite{Burda:2017tcu} found 
a variety of important special analytical cases that 
have been incorporated into {\tt 3VIL}.} using the differential equations method. 
The basis integral functions consist of a 1-loop integral
$A(x)$ already given above in eq.~(\ref{eq:defA}), the two-loop integral
$I(x,y,z)$ mentioned in the previous paragraph,
and three-loop integral functions $H(u,v,w,x,y,z)$, $G(w,u,z,v,y)$,
and $F(u,z,y,v)$, 
corresponding to topologies shown in Figure \ref{fig:basistopologies}.
For convenience, it is also useful to define a related function 
\beq
\Fbar(u,z,y,v) &=& F(u,z,y,v) + \lnbar(u) I(v,y,z)
\label{eq:defFbar}
\eeq
which is finite in the limit $u \rightarrow 0$, and an integral
$E(u,z,y,v)$ given by eq.~(2.40) in ref.~\cite{Martin:2016bgz}, which corresponds
to the same topology as $F(u,z,y,v)$ but without a derivative with respect to $u$.
In the following, I will employ $F(u,z,y,v)$ instead of $\Fbar(u,z,y,v)$ 
when the first argument does not vanish, and otherwise use $\Fbar(0,z,y,v)$.
Technically, $E(u,z,y,v)$ is not a basis integral because it can be written as a
linear combination of $F$ (or $\Fbar$) 
integrals with the same arguments in different orders, 
but it is convenient to use $E$ to express some quantities in simplest form.
Each of the basis integral functions is defined to include counterterms that 
make them finite and independent of the ultraviolet dimensional 
regularization parameter
$\epsilon$, but dependent on the \MSbar renormalization scale $Q$. 
This simplifies the presentation of results in the
\MSbar scheme, as $\epsilon$ never appears. 
Consult ref.~\cite{Martin:2016bgz} for the precise definitions 
of the basis integrals, and more information.

It is also convenient, when dealing with Feynman diagrams with ``doubled propagators" (i.e., two propagators that carry the same momentum)
to define functions by:
\beq
\Abar (w,x) &=& \left [ A(w) - A(x) \right ]/(x-w)
\qquad\qquad\qquad\qquad (x \not= w)
,
\\
\Ibar (w,x,y,z) &=& \left [ I(w,y,z) - I(x,y,z) \right ]/(x-w)
\qquad\qquad\qquad (x \not= w)
,
\\
K(w,x,u,z,y,v) &=& \left [ G(w,u,z,y,v) - G(x,u,z,y,v) \right ]/(x-w)
\qquad\quad (x \not= w)
.
\phantom{xxx}
\eeq
In each case, the first two arguments $w,x$ are the squared masses of the 
doubled propagators.
When the squared masses for double propagators coincide, these functions become:
\beq
\Abar(x,x) &=& -\frac{\partial}{\partial x} A(x) 
,
\label{eq:defAbarxx}
\\
\Ibar (x,x,y,z) &=& -\frac{\partial}{\partial x} I(x,y,z)
,
\\
K(x,x,u,z,y,v) &=&  -\frac{\partial}{\partial x} G(x,u,z,y,v)
,
\label{eq:defKxx}
\eeq
The necessary derivatives in eqs.~(\ref{eq:defAbarxx})-(\ref{eq:defKxx}) 
can be expressed in terms of the basis functions as:
\beq
\frac{\partial}{\partial x} A(x) &=& 1 + A(x)/x
,
\label{eq:dA}
\\
\frac{\partial}{\partial x} I(x,y,z) &=&
\bigl \{ (x-y-z) \left [I(x,y,z) - A(x) - A(y) - A(z) +x+y+z \right ]  
- 2 A(y) A(z) \phantom{xxx}
\nonumber \\ &&
+ (x -y + z) A(x) A(y)/x + (x+y-z) A(x) A(z)/x \bigr \}/\lambda(x,y,z)
,
\label{eq:dI}
\eeq
\vspace{-1.4cm}
\beq
\frac{\partial}{\partial x} G(x,u,z,y,v) &=&
\left [ (x-u-z)/\lambda(x,u,z) + (x-v-y)/\lambda(x,v,y) - 1/x \right ]
G(x,u,z,y,v)
\nonumber \\ &&
\!\!\!\!\!\!\!\!\!\!\!\!\!\!\!\!\!\!\!\!\!\!\!\!\!\!\!
+\bigl \{ 
[(x+u-z) A(z) + (x+z-u) A(u) + x (u+z-x)] I(x,v,y)
\nonumber \\ &&
\!\!\!\!\!\!\!\!\!\!\!\!\!\!\!\!\!\!\!\!\!\!\!\!\!\!\!
+ u (u-z-x)A(u)/4 + z (z-u-x)A(z)/4
+ x (u+z-x)[A(v) + A(y)] 
\nonumber \\ &&
\!\!\!\!\!\!\!\!\!\!\!\!\!\!\!\!\!\!\!\!\!\!\!\!\!\!\!
+ 2x[x^2 +x(2v + 2y - u - z) -2 u v - 2 u y -2 v z -2 y z - 8 u z]/3
\nonumber \\ &&
\!\!\!\!\!\!\!\!\!\!\!\!\!\!\!\!\!\!\!\!\!\!\!\!\!\!\!
+u (x+z-u) F(u,v,y,z) + z (x+u-z) F(z,u,v,y)
\bigr \}
/x\lambda(x,u,z) 
\nonumber \\ &&
\!\!\!\!\!\!\!\!\!\!\!\!\!\!\!\!\!\!\!\!\!\!\!\!\!\!\!
+\bigl \{ 
[(x+v-y) A(y) + (x+y-v) A(v) + x (v+y-x)] I(x,u,z)
\nonumber \\ &&
\!\!\!\!\!\!\!\!\!\!\!\!\!\!\!\!\!\!\!\!\!\!\!\!\!\!\!
+ v (v-y-x)A(v)/4 
+ y (y-v-x)A(y)/4
+ x (v+y-x)[A(u) + A(z)]
\nonumber \\ &&
\!\!\!\!\!\!\!\!\!\!\!\!\!\!\!\!\!\!\!\!\!\!\!\!\!\!\!
+ 2x[x^2 +x(2u + 2z - v - y) -2 u v - 2 u y -2 v z -2 y z - 8 v y]/3
\nonumber \\ &&
\!\!\!\!\!\!\!\!\!\!\!\!\!\!\!\!\!\!\!\!\!\!\!\!\!\!\!
+v (x+y-v) F(v,u,y,z) + y (x+v-y)  F(y,u,v,z) 
\bigr \}
/x\lambda(x,v,y) 
\nonumber \\ &&
\!\!\!\!\!\!\!\!\!\!\!\!\!\!\!\!\!\!\!\!\!\!\!\!\!\!\!
-(7x+2u+2v+2y+2z)/3x .
\label{eq:dG}
\eeq
Special cases that arise when the $\lambda(x,y,z)$
denominators vanish can be
obtained as smooth limits of the above. 
Of particular importance are the following:
\beq
\frac{\partial}{\partial x} I(x,0,x) &=& 
2 \left [\frac{\partial}{\partial x} I(x,0,z) \right ]\!\! \biggl |_{z=x}
\>=\> -A(x)^2/x^2,
\\
\frac{\partial}{\partial x} G(x,0,z,v,y) 
\Bigl |_{z=x}\!\! &=&
[F(x,0,v,y) - \Fbar(0,x,v,y)]/2x 
+ \bigl \{ 
[(v-y)^2 -x^2] A(x) I(x,v,y)/2x
\nonumber \\ &&
\!\!\!\!\!\!\!\!\!\!\!\!\!\!\!\!\!\!\!\!\!\!\!\!\!\!\!\!\!\!\!\!\!\!\!\!\!\!\!\!\!\!\!\!
+ [3x^2-4x(v+y) + (v-y)^2 ]I(x,v,y)/2
+ [(v-y-x) A(v) 
+ (y-v-x)A(y) 
\nonumber \\ &&
\!\!\!\!\!\!\!\!\!\!\!\!\!\!\!\!\!\!\!\!\!\!\!\!\!\!\!\!\!\!\!\!\!\!\!\!\!\!\!\!\!\!\!\!
+x(x-v-y)] A(x)^2/x
+ [2 A(v) A(y) 
+ 2 (x-y) A(y) 
+ 2 (x-v) A(v)
\nonumber \\ &&
\!\!\!\!\!\!\!\!\!\!\!\!\!\!\!\!\!\!\!\!\!\!\!\!\!\!\!\!\!\!\!\!\!\!\!\!\!\!\!\!\!\!\!\! 
+ (7v^2 + 7 y^2 + 18 v y 
+ 10 x v + 10 x y -17 x^2)/8] A(x) + x (v+y-x) [A(v) + A(y)]
\nonumber \\ &&
\!\!\!\!\!\!\!\!\!\!\!\!\!\!\!\!\!\!\!\!\!\!\!\!\!\!\!\!\!\!\!\!\!\!\!\!\!\!\!\!\!\!\!\! 
-2 x A(v) A(y) 
+ x[ 5 x^2 - 4 x (v+y) -v^2-y^2 - 10 v y]/3
\bigr \}/x\lambda(x,v,y) 
,
\eeq
and
\beq
\frac{\partial}{\partial x} G(x,0,y,0,y) \Bigl |_{y=x}&=&
1 -\Fbar(0,0,x,x)/x - A(x)/x
.
\eeq

It is often important to have expansions of the integral functions when one or more
of the squared mass arguments is small. In the following I will 
regulate infrared divergences in massless vector bosons by giving the propagator a
small squared mass (rather than using dimensional regularization for the infrared
divergences, which can cause confusion with the ultraviolet divergences). Goldstone
bosons also have squared masses that can be consistently 
treated as small compared to those of other particles.
The expansions of the basis integrals in a small squared mass $\delta$ 
(taking $\delta \ll x,y,z\ldots$, where $x,y,z\ldots$ are other pertinent
non-zero squared masses in the diagram)
can be accomplished using the differential equations that the basis integrals
satisfy, which were given in \cite{Martin:2016bgz}. As an example,
for the 2-loop basis integral function, one can find through order $\delta^2$ that,
for $x \not= y$: 
\beq
I(\delta,x,y) &=& I(0,x,y) +
\delta [-(x+y) I(0,x,y) - 2 A(x) A(y) + (3x-y) A(x) 
\nonumber \\ && 
+ (3y-x) A(y) - (x+y)^2]/(x-y)^2 + \delta \lnbar(\delta) \Abar(x,y) 
\nonumber \\ &&
+ \delta^2 [ -2 x y I(0,x,y) - (x+y) A(x) A(y) 
+ (7 x y - x^2 - 2 y^2) A(x)/2
\nonumber \\ &&
+ (7 x y - 2 x^2 - y^2) A(y)/2
+ (x+y) (2 x y - 5 x^2 - 5 y^2)/4 ]/(x-y)^4
\nonumber \\ &&
+ \delta^2 \lnbar(\delta) [ (x^2 - y^2)/2 + x A(y) - y A(x)]/(x-y)^3
+ {\cal O}(\delta^3) \qquad\qquad\qquad 
\eeq
and
\beq
I(\delta,x,x) &=& I(0,x,x) 
+ 
(\delta/x) [4 x + 3 A(x) + A(x)^2/2x -(x + A(x)) \lnbar(\delta) ]
\nonumber \\ &&
+
(\delta/x)^2 [-11 x/18 - A(x)/6 + x \lnbar(\delta)/6]
+ {\cal O}(\delta^3)
,
\\
I(\delta,0,x) &=& I(0,0,x) 
+ 
(\delta/x) [\zetatwo x + 2 A(x) + A(x)^2/2x - A(x) \lnbar(\delta) ]
\nonumber \\ &&
+
(\delta/x)^2 [-5 x/4 - A(x)/2 + x \lnbar(\delta)/2]
+ {\cal O}(\delta^3)
,
\\
I(\delta,\delta,x) &=& I(0,0,x) 
+ (\delta/x) [ 2 \zetatwo x + 4 A(x) + A(x)^2/x - 2 A(x) \lnbar(\delta)]
\nonumber \\ &&
+ (\delta/x)^2 [(2 \zetatwo  - 5/2)x + A(x) + A(x)^2/x - (x + 2 A(x)) \lnbar(\delta)
+ x \lnbar^2(\delta) ]
\nonumber \\ &&
+ {\cal O}(\delta^3)
,
\\
I(\delta,\delta,\delta) &=& 
\delta \left [3 \sqrt{3} {\rm Ls}_2 - 15/2 + 6 \lnbar(\delta) 
- 3 \lnbar^2(\delta)/2 \right ]
,
\\
I(0,\delta,\delta) &=& \delta \left[ -5 + 4 \lnbar(\delta) - \lnbar^2(\delta) \right ]
,
\\
I(0,0,\delta) &=& 
\delta \left[ -5/2 - \zetatwo + 2 \lnbar(\delta) - \lnbar^2(\delta)/2 \right ],
\\
I(0,0,0) &=& 0,
\eeq
where ${\rm Ls}_2 = -\int_0^{2\pi/3} dx \ln[2 \sin(x/2)] \approx 0.6766277376064358$.
A large number of similar expansion formulas for the 3-loop basis 
integrals, 
including all of the ones necessary for results below, are given
in an ancillary electronic file provided with this paper, called {\tt expzero.anc}.
In general, the functions $I$, $\Fbar$, $G$, and $H$ have smooth limits as 
$\delta \rightarrow 0$, with expansion terms that are powers of $\delta$ 
that may be multiplied by polynomials (of up to cubic order) in $\lnbar(\delta)$. 
The expansion of the function $F$ contains a $\lnbar(\delta)$ as the leading
behavior for $\delta \rightarrow 0$ 
if (and only if) the first argument
is $\delta$, as can be seen from eq.~(\ref{eq:defFbar}).
The limits for small $\delta$ of $\Abar(\delta,\delta)$ and
$\Ibar(\delta,\delta,x,y)$ and $K(\delta,\delta,u,v,x,y)$ have 
logarithmic infrared singularities,
because they also involve doubled propagators with the same momentum and the same
small squared mass $\delta$. 
Assuming that either $x$ or $y$ and either $u$ or $v$ are large compared
to $\delta$, one has:
\beq
\Abar(\delta,\delta) &=& -\lnbar(\delta)
,
\label{eq:Abardeldel}
\\
\Ibar(\delta,\delta,x,y) &=& -\lnbar(\delta) \Abar(x,y) + 
\ldots,
\\
K(\delta,\delta,u,v,x,y) 
&=& -\lnbar(\delta)  \Abar(u,v) \Abar(x,y)  + \ldots,
\label{eq:Kdeldel}
\eeq
where the ellipses refer to terms that are finite as $\delta \rightarrow 0$.
The expansions needed for the cases that occur in the 
Standard Model, through order $\delta^5$ for $I$, $F$ and
$\Fbar$ functions, through 
order $\delta^4$ for $\Ibar$ and $G$ functions, and through order $\delta^3$ for
$K$ and $H$ functions, are given in {\tt expzero.anc}. 
Further expansion cases as may be needed for more general theories can be obtained
by using the differential equations given in ref.~\cite{Martin:2016bgz}.

Finally, it is important to note that the loop integral basis functions satisfy
certain identities when the squared mass arguments are not generic, either because
some of them are equal to each other, or vanish. (These identities can be discovered
by requiring smooth limits of derivatives of the integral functions
as the arguments approach the
non-generic configurations.) Some identities of this type
were given in eqs.~(5.79)-(5.80) and (5.82)-(5.86) 
of ref.~\cite{Martin:2016bgz}. Other identities
that are used in the following are:
\beq
F(x,x,y,y) &=& (x/y-1) [F(x,0,0,y) + I(0,x,y) + A(y) - 2 y \zeta_3] 
\nonumber \\ &&
+ [A(y)/y - A(x)/x] I(0,x,y)
+ A(x) [A(y)^2/y + A(x) - 2 A(y) 
\nonumber \\ &&
+ 3 x^2/4y - 9 x/2 + 2 y]/x
- 2 x^2/3y + 10 x/3 + 2 y
,
\\
G(0,0,0,x,y) &=& -\Fbar(0,0,x,y) + 2 I(0,x,y) + A(x) + A(y)
-4x/3 - 4y/3 ,
\\
G(y,0,0,0,x) &=& G(x,0,0,0,y) + 
(x-y) [F(x,0,0,y) + I(0,x,y) + A(x) A(y)/x 
\nonumber \\ &&
+ 2 y \zetathree - 2 (x+5y)/3]/y + A(x) A(y) [A(y) - A(x)]/(2 x y)
\nonumber \\ &&
+ [1/4 + 3x/4y + (1 + \zetatwo) y/x] A(x)
- [2 + x\zetatwo /y] A(y) 
,
\\
G(x,0,y,0,y) &=& (y-x) [F(x,0,0,y)/y + \Fbar(0,0,x,y)/x] 
+ 2 A(x) A(y)/x 
\nonumber \\ &&
+ [3 - x/y - (1/x+1/y) A(y)] I(0,x,y)
+ [y - A(x) - A(y)/3] A(y)^2/x y
\phantom{xxx}
\nonumber \\ &&
+ [(3 x y/4 -y^2 - 3 x^2/4) A(x) - (x-y)^2 A(y)]/x y
\nonumber \\ &&
+ 2 (x^2 + 2 x y - 7 y^2)/3y
+ 2 (y^2 + 6 x y  - 3 x^2) \zetathree/3x 
.
\eeq
In addition, one can express the following 1-scale integrals in terms of 
$\zetatwo$ and $\zetathree$ and powers of $A(x)$, with rational coefficients, 
using the analytical formulas collected
in section V of ref.~\cite{Martin:2016bgz}:
$I(0,0,x)$, $I(0,x,x)$, $F(x,0,0,0)$, $F(x,0,0,x)$, $F(x,x,x,x)$, 
$\Fbar(0,0,0,x)$, $\Fbar(0,0,x,x)$, $G(0,0,0,0,x)$, 
$G(0,0,0,x,x)$, $G(0,0,x,0,x)$, 
$G(x,0,0,0,0)$,
$G(x,0,0,0,x)$, $G(x,0,x,0,x)$, and $G(0,x,x,x,x)$. 
The existence of these identities means that the presentation of 
results for any specific theory (for example, the Standard Model)
in terms of the basis functions is far from unique; 
the basis is over-complete when the arguments are not generic.

\section{Three-loop contributions to the effective potential\label{sec:mainresults}}
\setcounter{equation}{0}
\setcounter{figure}{0}
\setcounter{table}{0}
\setcounter{footnote}{1}

\subsection{Feynman diagrams\label{subsec:Feynmandiagrams}}

\begin{figure}[t]
\begin{center}
\includegraphics[width=0.99\linewidth,angle=0]{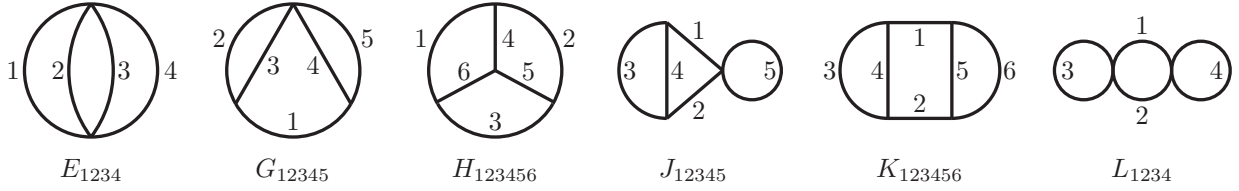}
\end{center}
\vspace{-0.25cm}
\begin{minipage}[]{0.96\linewidth}
\caption{\label{fig:diagramtopologies}
The Feynman diagram topologies contributing to the 3-loop effective potential,
with numerals indicating the ordering of subscripts denoting propagator
types ($S$, $F$, $\Fbar$, $V$, or $g$) as well as the ordering of the
corresponding squared mass arguments.}
\end{minipage}
\end{figure}

\begin{figure}[t]
\phantom{x}
\begin{center}
\includegraphics[width=0.56\linewidth,angle=0]{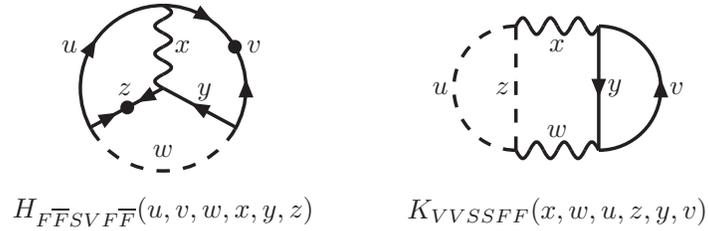}
\end{center}
\vspace{-0.25cm}
\begin{minipage}[]{0.96\linewidth}
\caption{\label{fig:diagramexamples}
Examples of the Feynman diagram labeling scheme used in this paper, for the
diagrams with loop integral functions denoted $H_{F\Fbar SVF\Fbar}(u,v,w,x,y,z)$
and $K_{VVSSFF}(x,w,u,z,y,v)$. Solid lines with arrows represent helicity-preserving
fermion propagators. Solid lines with a dot and clashing arrows represent a helicity-violating fermion propagator. Dashed lines indicate a real scalar propagator,
and wavy lines stand for real vector propagators. The squared masses are denoted by
$u,v,w,x,y,z$ as labeled.}
\end{minipage}
\end{figure}

In this section I present the results for the three-loop contribution to the effective
potential for a general renormalizable quantum field theory. 
The 1-particle-irreducible Feynman diagrams for the three-loop 
effective potential have the topologies shown in
Figure \ref{fig:diagramtopologies}.
To distinguish the different diagrams, 
each topology is associated with a
letter $E$, $G$, $H$, $J$, $K$, or $L$, and then
subscripts $S$, $V$, $F$, $\Fbar$, or $g$ are applied,
corresponding respectively to real scalar, real vector,  
helicity-preserving fermion, helicity-violating fermion,
or ghost propagators, in the order designated by the numbering in 
Figure~(\ref{fig:diagramtopologies}). 
The helicity-violating fermion propagators each contain a mass insertion of
the type $M_{II'}$ or $M^{II'}$, 
as described in ref.~\cite{Dreiner:2008tw}. To illustrate this
labeling scheme, Figure \ref{fig:diagramexamples} shows the Feynman diagrams
corresponding to diagrams
$H_{F\Fbar SV F\Fbar}$ and $K_{VVSSFF}$.
For each diagram, there is a corresponding loop integral function,
which one can compute in terms of the basis functions discussed in the previous
section after including the \MSbar counterterms. The squared mass arguments are 
given in the same ordering as the corresponding subscripts.

However, the correspondence between Feynman diagrams and loop integral functions is
not one-to-one, because in some cases involving vector bosons
it is convenient to define loop integral functions that
combine the effects of more than one Feynman diagram, by exploiting the constraints
implied by the underlying gauge invariance that is associated with vector fields in
renormalizable theories. For example, 
because of the relation between the vector-scalar-scalar couplings 
$g^{ajk}$ and the vector-vector-scalar-scalar couplings 
$g^{abjk}$ given in eq.~(\ref{eq:VVSScouplings}), 
it is convenient to define a single function $K_{SSSSSV}(u,v,w,x,y,z)$ 
that combines the effect of
the Feynman diagram labeled $K_{SSSSSV}$ with the one labeled $J_{SSSSV}$. 
(For this reason, there is no function $J_{SSSSV}$ below.)
There are numerous similar cases where the contribution
of a diagram with a vector-vector-scalar-scalar interaction 
is combined with the contribution from
a related diagram with a pair of vector-scalar-scalar interactions to give a single 
loop integral function.
Furthermore, because of 
the fact that the vector quartic interaction and the vector-ghost-antighost
interaction are determined by the triple vector coupling, as seen in 
eq.~(\ref{eq:Linteractions}),
the effects of the diagrams
labeled $H_{VVVVVV}$, $H_{VgggVg}$, $H_{gggVVV}$, and parts of $G_{VVVVV}$ and $E_{VVVV}$
can always be combined into a single function that I  
call $H_{\mbox{gauge}}$.
The other parts of diagrams $G_{VVVVV}$ and $E_{VVVV}$, together with the contributions
of diagrams
$K_{VVVVVV}$, $K_{VVgggg}$, $K_{VVVVgg}$, $K_{gggVVg}$, $J_{VVVVV}$, $J_{VVggV}$,
and $L_{VVVV}$ can always 
be combined into a single function to be denoted $K_{\mbox{gauge}}$. 
Similarly, I define a function $H_{\mbox{gauge},S}$ that combines
the effects from the diagrams $H_{VVVVVS}$ and $G_{SVVVV}$; a function
$K_{\mbox{gauge},S}$ that combines the effects of diagrams $K_{VVSVVV}$,
$K_{VVSVgg}$, and $J_{VVVSV}$; a function $K_{\mbox{gauge},SS}$ that combines the
contributions from $K_{VVSSVV}$, $K_{VVSSgg}$, $J_{VVSSV}$, $J_{VVVVS}$, $J_{VVggS}$, and
$L_{VVVS}$; a function
$K_{\mbox{gauge},FF}$ that combines diagrams $K_{VVVVFF}$, $K_{VVFFgg}$, and $J_{VVFFV}$;
and a function $K_{\mbox{gauge},\Fbar\Fbar}$ that combines diagrams 
$K_{VVVV\Fbar\Fbar}$, $K_{VV\Fbar\Fbar gg}$, and $J_{VV\Fbar\Fbar V}$.

Finally, note that there are two diagrams, $G_{VSSSS}$ and $G_{VSSVV}$, 
which one can draw and for which the couplings exist, but for which the corresponding 
loop integrals vanish identically. 

Taking into account the above considerations, I find that the three-loop
contributions to the \MSbar renormalized Landau gauge effective potential
for a general renormalizable theory can be expressed in terms of 
only 89 distinct loop integral functions:
\beq
&& \!\!\!\!\!\!\!
H_{SSSSSS},\> 
K_{SSSSSS},\>
J_{SSSSS},\>
G_{SSSSS},\> 
L_{SSSS},\> 
E_{SSSS},\> 
H_{FF\Fbar SSS},\>
H_{\Fbar\Fbar\Fbar SSS},\>
H_{FFSSFF},\>\phantom{xx}
\nonumber \\ && \!\!\!\!\!\!\!
H_{FFSS\Fbar\Fbar},\>
H_{F\Fbar SSF\Fbar},\>
H_{\Fbar\Fbar SS\Fbar\Fbar},\>
K_{SSSSFF},\>
K_{SSSS\Fbar\Fbar},\>
K_{FFFSSF},\>                                                                                                                     
K_{FF\Fbar SS\Fbar},\>
\nonumber \\ && \!\!\!\!\!\!\!
K_{\Fbar\Fbar FSSF},\>      
K_{\Fbar F\Fbar SSF},\>                                                                            
K_{\Fbar\Fbar\Fbar SS\Fbar},\>                                                                           
K_{SSFFFF},\>
K_{SSFF\Fbar\Fbar},\>
K_{SS\Fbar\Fbar\Fbar\Fbar},\>
J_{SSFFS},\> 
J_{SS\Fbar\Fbar S},\>     
\nonumber \\ && \!\!\!\!\!\!\!
H_{SSSSSV},\>
H_{VVSSSS},\>
H_{SSVVSS},\>
H_{VVVSSS},\>
H_{SSSVVV},\>
H_{VVSSVS},\>
H_{SSVVVV},\>
\nonumber \\ && \!\!\!\!\!\!\!
H_{SVVVSV},\>
K_{SSSSSV},\>
K_{SSSSVV},\>
K_{SSSVVS},\>
K_{VVSSSS},\>
K_{SSSVVV},\>
K_{VVSSVS},\>
\nonumber \\ && \!\!\!\!\!\!\!
K_{SSVVVV},\>
K_{VVSVVS},\>
J_{SSVSS},\>
J_{SSVVS},\>
G_{VSVVS},\>
H_{\mbox{gauge},\>S},\>
K_{\mbox{gauge},\>S},\>
K_{\mbox{gauge},\>SS},\>
\nonumber \\ && \!\!\!\!\!\!\!
H_{FFVVFF},\>
H_{FFVV\Fbar\Fbar},\>
H_{F\Fbar VVF\Fbar},\>
H_{\Fbar\Fbar VV\Fbar\Fbar},\>
H_{FFFVVV},\>
H_{F\Fbar\Fbar VVV},\>
K_{FFFVVF},\>
\nonumber \\ && \!\!\!\!\!\!\!
K_{FF\Fbar VV\Fbar},\>
K_{\Fbar\Fbar FVVF},\>
K_{\Fbar F\Fbar VVF},\>
K_{\Fbar\Fbar\Fbar VV\Fbar},\>
K_{VVFFFF},\>
K_{VVFF\Fbar\Fbar},\>
K_{VV\Fbar\Fbar\Fbar\Fbar},\>
\nonumber \\ && \!\!\!\!\!\!\!
K_{\mbox{gauge},\>FF},\>
K_{\mbox{gauge},\>\Fbar\Fbar},\>
H_{FFSVFF},\>
H_{FFSV\Fbar\Fbar},\>
H_{F\Fbar SV\Fbar F},\>
H_{F\Fbar SVF\Fbar},\>
H_{\Fbar\Fbar SV\Fbar\Fbar},\>
\nonumber \\ && \!\!\!\!\!\!\!
H_{FFFVSS},\>
H_{F\Fbar\Fbar VSS},\>
H_{\Fbar\Fbar FVSS},\>
H_{F\Fbar FSVV},\>
H_{FF\Fbar SVV},\>
H_{\Fbar\Fbar\Fbar SVV},\>
K_{FFFSVF},\>
\nonumber \\ &&\!\!\!\!\!\!\!
K_{FF\Fbar SV\Fbar},\>
K_{\Fbar\Fbar FSVF},\>
K_{\Fbar F\Fbar SVF},\>
K_{\Fbar FF SV\Fbar},\>
K_{\Fbar\Fbar\Fbar SV\Fbar},\>
K_{SSSVFF},\>
K_{SSSV\Fbar\Fbar},\>
\nonumber \\ &&\!\!\!\!\!\!\!
K_{SSVVFF},\>
K_{SSVV\Fbar\Fbar},\>
K_{VVSSFF},\>
K_{VVSS\Fbar\Fbar},\>
K_{VVSVFF},\>
K_{VVSV\Fbar\Fbar},\>
H_{\mbox{gauge}},\>
K_{\mbox{gauge}}.
\label{eq:genloopfunctions}
\eeq
It remains to give $V^{(3)}$ by 
providing expressions for these 89 functions in terms of the basis
functions described in the previous section, with arguments that are \MSbar
squared masses (and, implicitly, the renormalization scale $Q$), and also to
provide the coefficients of these 89 functions in $V^{(3)}$ 
in terms of the \MSbar 
couplings appearing in eq.~(\ref{eq:Linteractions}). 
Regarding the first task,
many of the expressions for the 89 loop integral functions in terms of
basis integrals are extremely complicated and not of much use to the human eye. 
All of these results are therefore presented in an ancillary electronic file
called {\tt functions.anc} distributed with this paper, suitable for inclusion in
symbolic manipulation code or numerical computer programs.
Only the first 24, relatively simple, functions in eq.~(\ref{eq:genloopfunctions}),
corresponding to the diagrams that do not involve vector propagators 
will be given in the text below.

The formula for $V^{(3)}$ in terms of the 89 functions listed in 
eq.~(\ref{eq:genloopfunctions}) is split up below as:
\beq
V^{(3)} = V^{(3)}_S + V^{(3)}_{SF} + V^{(3)}_{SV} 
+ V^{(3)}_{FV} + V^{(3)}_{SFV} + V^{(3)}_V,
\eeq
where 
$V^{(3)}_S$ contains only scalar interactions, 
$V^{(3)}_{SF}$ contains scalars and fermions only, 
$V^{(3)}_{SV}$ contains only scalars and vectors (and ghosts), 
$V^{(3)}_{FV}$ contains only fermions and vectors (and ghosts),
$V^{(3)}_{SFV}$ contains scalars, fermions, and vectors (and ghosts), and
$V^{(3)}_{V}$ contains only vectors and ghosts.

\subsection{Pure scalar contributions\label{subsec:V3S}}

The pure scalar contributions to the three-loop effective potential can be written
as:
\beq
V^{(3)}_S &=& 
     \frac{1}{24} \lambda^{jkm} \lambda^{kln} \lambda^{jlp} \lambda^{mnp}
     H_{SSSSSS}(j,k,l,m,n,p)
\nonumber \\ &&        
   + \frac{1}{16} \lambda^{jlm} \lambda^{klm} \lambda^{jnp} \lambda^{knp}
     K_{SSSSSS}(j,k,l,m,n,p) 
\nonumber \\ &&   
   + \frac{1}{8} \lambda^{jknn} \lambda^{jlm} \lambda^{klm}
     J_{SSSSS}(j,k,l,m,n)
   + \frac{1}{8} \lambda^{jkl} \lambda^{jmn} \lambda^{klmn} 
     G_{SSSSS}(j,k,l,m,n) 
\nonumber \\ &&   
   + \frac{1}{16} \lambda^{jkll} \lambda^{jkmm} 
     L_{SSSS}(j,k,l,m)
   + \frac{1}{48} \lambda^{jklm} \lambda^{jklm} 
     E_{SSSS}(j,k,l,m) ,
\label{eq:V3loopSgeneral}
\eeq
where the scalar field indices $j,k,l,m,n,p$ are also used to represent the
\MSbar
background-field-dependent squared masses.
The loop integral functions appearing in eq.~(\ref{eq:V3loopSgeneral}) 
are easy to write in terms of the basis integrals:
\beq
H_{SSSSSS}(u,v,w,x,y,z) &=& -H(u,v,w,x,y,z)
,
\label{eq:defHSSSSSS}
\\
K_{SSSSSS}(u,v,w,x,y,z) &=& -K(u,v,w,x,y,z)
,
\\
J_{SSSSS}(w,x,v,y,u) &=& A(u) \Ibar(w,x,v,y)
,
\\
G_{SSSSS}(w,u,z,y,v) &=& G(w,u,z,y,v) 
,
\\
L_{SSSS}(w,x,u,v) &=& -A(u) A(v) \Abar(w,x)
,
\\
E_{SSSS}(u,z,y,v) &=& -E(u,z,y,v)
.
\label{eq:defESSSS}
\eeq

\subsection{Scalar and fermion contributions\label{subsec:V3SF}}

The contributions involving scalars and fermions (but not vectors or ghosts) can be
written as:
\beq
V^{(3)}_{SF} &=&  
    \frac{1}{2}  \left(\lambda^{jkl} Y^{jIJ} Y_{kJK} Y_{lIK'} M^{KK'} 
    + \cc \right ) 
    H_{FF\Fbar SSS}(I,J,K,j,k,l)
\nonumber \\ &&   
    + \frac{1}{6} \left(\lambda^{jkl} Y^{jIJ} Y^{kJ'K} Y^{lI'K'} 
    M_{II'} M_{JJ'} M_{KK'} +\cc \right ) 
    H_{\Fbar\Fbar\Fbar SSS}(I,J,K,j,k,l)
\nonumber \\ &&   
    + \frac{1}{4} Y^{kIJ} Y^{kKL} Y_{jIL} Y_{jJK}
    H_{FFSSFF}(I,J,j,k,K,L)  
\nonumber \\ &&   
    + \frac{1}{2} \left (Y^{kIJ} Y_{kKL} Y_{jIL'} Y_{jJK'} M^{KK'} M^{LL'} 
    + \cc \right )
    H_{FFSS\Fbar\Fbar}(I,J,j,k,K,L)
\nonumber \\ &&   
    + \frac{1}{2} Y^{kIJ} Y_{kKL} Y_{jIL'} Y^{jJ'K} M_{JJ'} M^{LL'} 
    H_{F\Fbar SSF\Fbar}(I,J,j,k,K,L) 
\nonumber \\ &&   
    + \frac{1}{8} \left (Y^{kIJ} Y^{kKL} Y^{jI'L'} Y^{jJ'K'} 
    M_{II'} M_{JJ'} M_{KK'} M_{LL'} + \cc \right )
    H_{\Fbar\Fbar SS\Fbar\Fbar}(I,J,j,k,K,L)  
\nonumber \\ &&   
    + \frac{1}{4} \lambda^{jlm} \lambda^{klm} Y^{jIJ} Y_{kIJ}
    K_{SSSSFF}(j,k,l,m,I,J) 
\nonumber \\ &&   
    + \frac{1}{8} \left(\lambda^{jlm} \lambda^{klm}  Y^{jIJ} Y^{kI'J'} M_{II'} M_{JJ'} 
    + \cc \right )
    K_{SSSS\Fbar\Fbar}(j,k,l,m,I,J) 
\nonumber \\ &&   
    + \frac{1}{2} Y^{jIK} Y_{jJK} Y_{kIL} Y^{kJL}
    K_{FFFSSF}(I,J,K,j,k,L)                                                                                                                     
\nonumber \\ &&   
    + \frac{1}{2} Y^{jIK} Y^{jJK'} Y_{kIL} Y_{kJL'} M_{KK'} M^{LL'}
    K_{FF\Fbar SS\Fbar}(I,J,K,j,k,L)                                       
\nonumber \\ &&   
    + \frac{1}{2} Y^{jIK} Y_{jJK} Y^{kI'L} Y_{kJ'L} M_{II'} M^{JJ'}
    K_{\Fbar\Fbar FSSF}(I,J,K,j,k,L)      
\nonumber \\ &&   
    + \left( Y^{jIK} Y^{jJK'} Y^{kI'L} Y_{kJL} M_{II'} M_{KK'} + \cc \right )
    K_{\Fbar F\Fbar SSF}(I,J,K,j,k,L)                                                                            
\nonumber \\ &&   
    + \frac{1}{4} \left( Y^{jIK} Y^{jJK'} Y^{kI'L} Y^{kJ'L'} 
    M_{II'} M_{JJ'} M_{KK'} M_{LL'} + \cc \right )
    K_{\Fbar\Fbar\Fbar SS\Fbar}(I,J,K,j,k,L)                                                                            
\nonumber \\ &&   
    + \frac{1}{8} Y^{jIJ} Y_{kIJ} \left (Y^{jKL}Y_{kKL} + \cc\right )
    K_{SSFFFF}(j,k,I,J,K,L)                                                                                                                     
\nonumber \\ &&   
    + \frac{1}{4} Y^{jIJ} Y_{kIJ} \left (Y^{jKL}Y^{kK'L'}M_{KK'} M_{LL'} + \cc\right )
    K_{SSFF\Fbar\Fbar}(j,k,I,J,K,L)                                                                                                                     
\nonumber \\ &&   
    + \frac{1}{16} \left(Y^{jIJ} Y^{kI'J'} M_{II'} M_{JJ'} +\cc \right ) 
    \left (Y^{jKL}Y^{kK'L'}M_{KK'} M_{LL'} + \cc\right )
    K_{SS\Fbar\Fbar\Fbar\Fbar}(j,k,I,J,K,L)  
\nonumber \\ &&   
    + \frac{1}{4} \lambda^{jkll} Y^{jIJ} Y_{kIJ}     
    J_{SSFFS}(j,k,I,J,l)  
\nonumber \\ &&   
    + \frac{1}{8}  \left (\lambda^{jkll} Y^{jIJ} Y^{kI'J'} M_{II'} M_{JJ'}     
    + \cc \right )
    J_{SS\Fbar\Fbar S}(j,k,I,J,l)   .                                                                                                                                                                                                                                                 
\label{eq:V3loopSFgeneral}
\eeq
I find that the loop integral functions appearing in eq.~(\ref{eq:V3loopSFgeneral}) are,
in terms of the basis integrals:
\beq
H_{FF\Fbar SSS}(u,v,w,x,y,z) &=& (u + v - x) H(u,v,w,x,y,z) 
+ G(w, u, z, v, y) 
\nonumber \\ &&
- G(y, v, w, x, z) 
- G(z, u, w, x, y)
,
\label{eq:defHFFfSSS}
\\
H_{\Fbar\Fbar\Fbar SSS}(u,v,w,x,y,z) &=& 2 H(u,v,w,x,y,z)   
,
\\
H_{FFSSFF}(u,v,w,x,y,z) &=& 
(u y + v z - w x) H(u, v, w, x, y, z) 
- u G(u, v, x, w, z) 
\nonumber \\ &&
- v G(v, u, x, w, y) 
+ w G(w, u, z, v, y) 
+ x G(x, u, v, y, z) 
\nonumber \\ &&
- y G(y, v, w, x, z) 
- z G(z, u, w, x, y)  
- E(u,v,y,z) 
\nonumber \\ &&
+ E(v,w,x,z) + E(u,w,x,y)
,
\\
H_{FFSS\Fbar\Fbar}(u,v,w,x,y,z) &=& 
(u + v - x) H(u, v, w, x, y, z) 
+ G(w, u, z, v, y) 
\nonumber \\ &&
- G(y, v, w, x, z) 
- G(z, u, w, x, y)
,
\\
H_{F\Fbar SSF\Fbar}(u,v,w,x,y,z) &=& 
(v - w - x + z) H(u, v, w, x, y, z)
- G(v, u, x, w, y) 
\nonumber \\ &&  
+ G(w, u, z, v, y) 
+ G(x, u, v, y, z) 
- G(z, u, w, x, y) 
,
\\
H_{\Fbar\Fbar SS\Fbar\Fbar}(u,v,w,x,y,z) &=& 2 H(u,v,w,x,y,z)   
,
\\
K_{SSSSFF}(x,w,u,z,y,v) &=&
[v+y -(w+x)/2] K(x,w,u,z,y,v)
+ G(w,u,z,y,v)/2 
\nonumber \\ &&  
+ G(x,u,z,y,v)/2 
- [A(v) + A(y)] \Ibar(x,w,u,z)
,
\\
K_{SSSS\Fbar\Fbar}(x,w,u,z,y,v) &=& 2 K(x,w,u,z,y,v)
,
\\
K_{FFFSSF}(x,w,u,z,y,v) &=& 
(x^2 + w^2 + 2 u v - 2 u y - 2 v z + 2 y z 
+ u w + v w  
\nonumber \\ && 
+ u x + v x - w y - x y  - w z - x z ) K(x,w,u,z,y,v)/4
\nonumber \\ &&
+(y + z - u - v - w - x)[G(w,u,z,y,v) + G(x,u,z,y,v)]/4
\nonumber \\ &&
+(u+w-z) [A(v) - A(y)] \Ibar(x,w,u,z)/2
\nonumber \\ &&
+ (v+w-y) [A(u) - A(z)] \Ibar(x,w,v,y)/2
+ E(u,v,y,z)/2
\nonumber \\ &&
+ [A(y) - A(v)] I(u,x,z)/2
+ [A(z) - A(u)] I(v,x,y)/2
\nonumber \\ &&
+ \Abar(x,w) [A(z) - A(u)][A(y) - A(v)]/2 
,
\\
K_{FF\Fbar SS\Fbar}(x,w,u,z,y,v) &=& 
(w+x) K(x,w,u,z,y,v) - G(w,u,z,y,v) - G(x,u,z,y,v)
,\phantom{xxxx}
\\
K_{\Fbar\Fbar FSSF}(x,w,u,z,y,v) &=& \Bigl \{
(w^2 x + w x^2 + 2 u w x + 2 v w x - 2 w x y + u v w + u v x - u w y 
\nonumber \\ &&
- u x y  - v w z - v x z - 2 w x z + w y z + x y z) K(x,w,u,z,y,v)
\nonumber \\ &&
+ (u v - w x - u y - v z + y z) [G(x,u,z,y,v) + G(w,u,z,y,v)]
\nonumber \\ &&
+ 2 (v-y) [u F(u,v,y,z) - z F(z,u,v,y) + [A(u) - A(z)] I(v,x,y)] 
\nonumber \\ &&
+ 2 (u-z) [v F(v,u,y,z) - y F(y,u,v,z) + [A(v) - A(y)] I(u,x,z)]
\nonumber \\ &&
+ 2 x (u+w-z) [A(v) - A(y)] \Ibar(w,x,u,z)
\nonumber \\ &&
+ 2 x (v+w-y) [A(u) - A(z)] \Ibar(w,x,v,y)
\nonumber \\ &&
+ 2 [x \Abar(w,x) + A(x)][A(v) - A(y)] [A(u) - A(z)] 
\nonumber \\ &&
+ (u-z) [y A(y) - v A(v)]/2 
+ (v-y) [z A(z) - u A(u)]/2
\nonumber \\ &&
+ 4 (v - y) (u - z) (u + v + y + z)/3
\Bigr \}/4wx
,
\label{eq:KffFSSF}
\\
K_{\Fbar F\Fbar SSF}(x,w,u,z,y,v) &=& [v-y + (w+x)/2] K(x,w,u,z,y,v) 
-G(w, u, z, v, y)/2 
\nonumber \\ &&
- G(x, u, z, v, y)/2 
+ [A(v) - A(y)] \Ibar(x,w,u,z)
,
\\
K_{\Fbar\Fbar\Fbar SS\Fbar}(x,w,u,z,y,v) &=& 2 K(x,w,u,z,y,v)
,
\\
K_{SSFFFF}(x,w,u,z,y,v) &=&
[(w + x) (u + v + y + z) -2(u+z)(v+y) 
\nonumber \\ &&
-w^2 -x^2]K(x,w,u,z,y,v)/2 - E(u,v,y,z)
\nonumber \\ &&
+ (w+x-u-v-y-z) [G(w,u,z,y,v) + G(x,u,z,y,v)]/2
\nonumber \\ &&
+ (u-w+z) [A(v) + A(y)] \Ibar(w,x,u,z)
\nonumber \\ &&
+ (v-w+y) [A(u) + A(z)] \Ibar(w,x,v,y)
\nonumber \\ &&
+[A(v) + A(y)] I(u,x,z)
+[A(u)] + A(z)] I(v,x,y)
\nonumber \\ &&
- \Abar(w,x) [A(u) + A(z)][A(v) + A(y)]  
,
\\
K_{SSFF\Fbar\Fbar}(x,w,u,z,y,v) &=&
(w+x-2u-2z) K(x,w,u,z,y,v)
- G(w, u, z, v, y) 
\nonumber \\ &&
- G(x, u, z, v, y) 
+ 2 [A(u) + A(z)] \Ibar(x,w,y,v)
,
\\
K_{SS\Fbar\Fbar\Fbar\Fbar}(x,w,u,z,y,v) &=& -4 K(x,w,u,z,y,v)
,
\\
J_{SSFFS}(x,w,y,v,u) &=& A(u) \bigl \{(w-v-y) \Ibar(x,w,y,v) - I(v,x,y) 
\nonumber \\ &&
+ \Abar(x,w) [A(v) + A(y)] \bigr \}
,
\\
J_{SS\Fbar\Fbar S}(x,w,y,v,u) &=& -2 A(u) \Ibar(x,w,y,v)
.
\label{eq:defJSSffS}
\eeq
These can also be found in the ancillary electronic file {\tt functions.anc}.

\subsection{Scalar and vector (and ghost) contributions\label{subsec:V3SV}}

The contributions that involve both scalars and vectors, but not fermions, are written as:
\beq
V^{(3)}_{SV} &=&   
     \frac{1}{4} \lambda^{jkm} \lambda^{kln} g^{alj} g^{amn}
     H_{SSSSSV}(j,k,l,m,n,a)
\nonumber \\ &&   
   + \frac{1}{2} \lambda^{klm} g^{abk} g^{ajm} g^{blj}
     H_{VVSSSS}(a,b,j,k,l,m)
\nonumber \\ &&   
   + \frac{1}{8} g^{ajm} g^{bjk} g^{akl} g^{blm}
     H_{SSVVSS}(j,k,a,b,l,m)     
\nonumber \\ &&   
   + \frac{1}{6} \lambda^{jkl} g^{abj} g^{bck} g^{acl}  
     H_{VVVSSS}(a,b,c,j,k,l)     
\nonumber \\ &&   
   + \frac{1}{6} g^{abc} g^{ajk} g^{bkl} g^{clj} 
     H_{SSSVVV}(j,k,l,a,b,c)    
\nonumber \\ &&   
   + \frac{1}{2} g^{abk} g^{bcj} g^{ajl} g^{clk} 
     H_{VVSSVS}(a,b,j,k,c,l)         
\nonumber \\ &&   
   + \frac{1}{2} g^{ack} g^{adj} g^{bcd} g^{bjk} 
     H_{SSVVVV}(j,k,a,b,c,d)     
\nonumber \\ &&   
   + \frac{1}{8} g^{acj} g^{abk} g^{bdj} g^{cdk} 
     H_{SVVVSV}(j,a,b,c,k,d) 
\nonumber \\ &&   
   + \frac{1}{4} \lambda^{jlm} \lambda^{klm} g^{ajn} g^{ank}
     K_{SSSSSV}(j,k,l,m,n,a)
\nonumber \\ &&   
   + \frac{1}{8} \lambda^{jlm} \lambda^{klm} g^{abj} g^{abk}
     K_{SSSSVV}(j,k,l,m,a,b)
\nonumber \\ &&   
   + \frac{1}{4} g^{ajl} g^{alk} g^{bjm} g^{bmk} 
     K_{SSSVVS}(j,k,l,a,b,m)
\nonumber \\ &&   
   + \frac{1}{16} g^{ajk} g^{bkj} g^{aml} g^{blm}
     K_{VVSSSS}(a,b,j,k,l,m)
\nonumber \\ &&   
   + \frac{1}{4} g^{ajl} g^{alk} g^{bcj} g^{bck}
     K_{SSSVVV}(j,k,l,a,b,c)
\nonumber \\ &&   
   + \frac{1}{4} g^{ajk} g^{bkj} g^{acl} g^{bcl}
     K_{VVSSVS}(a,b,j,k,c,l)
\nonumber \\ &&   
   + \frac{1}{16} g^{abj} g^{abk} g^{cdj} g^{cdk}
     K_{SSVVVV}(j,k,a,b,c,d)
\nonumber \\ &&   
   + \frac{1}{4} g^{acj} g^{bcj} g^{adk} g^{bdk}
     K_{VVSVVS}(a,b,j,c,d,k)
\nonumber \\ &&   
   + \frac{1}{4} \lambda^{jkmm} g^{ajl} g^{alk}
     J_{SSVSS}(j,k,a,l,m)  
\nonumber \\ &&   
   + \frac{1}{8} g^{abj} g^{abk} \lambda^{jkll}
     J_{SSVVS}(j,k,a,b,l)     
\nonumber \\ &&   
   + \frac{1}{2} g^{abj} g^{ack} g^{bjl} g^{ckl}  
     G_{VSVVS}(a,j,b,c,k)     
\nonumber \\ &&  
   + \frac{1}{4} g^{abd} g^{bce} g^{acj} g^{dej} 
   H_{\mbox{gauge},\>S}(a,b,c,d,e,j)   
\nonumber \\ &&  
   + \frac{1}{4} g^{acd} g^{bcd} g^{aej} g^{bej} 
   K_{\mbox{gauge},\>S}(a,b,c,d,e,j)                                                               
\nonumber \\ &&  
   + \frac{1}{8} g^{acd} g^{bcd} g^{ajk} g^{bkj} 
   K_{\mbox{gauge},\>SS}(a,b,c,d,j,k)                                                               
.
\eeq
The loop integral functions appearing here are presented explicitly in the ancillary
electronic file {\tt functions.anc}, in computer-readable form. 
Many of them are quite lengthy. 

\subsection{Fermion and vector (and ghost) contributions\label{subsec:V3FV}}

The contributions involving fermions and vectors (but not scalars) are written as:
\beq
V^{(3)}_{FV} &=& 
   \frac{1}{4} g^{aL}_I g^{bK}_L g^{aJ}_K g^{bI}_J
   H_{FFVVFF}(I,J,a,b,K,L)
\nonumber \\ &&  
   + g^{aL}_I g^{bL'}_{K} g^{aJ}_{K'} g^{bI}_J M^{KK'} M_{LL'}
   H_{FFVV\Fbar\Fbar}(I,J,a,b,K,L)
\nonumber \\ &&  
   + \frac{1}{2} g^{aL}_I g^{bL'}_{K} g^{aK}_{J} g^{bI}_{J'} M^{JJ'} M_{LL'}
   H_{F\Fbar VVF\Fbar}(I,J,a,b,K,L)
\nonumber \\ &&  
   + \frac{1}{4} g^{aI}_L g^{bK}_{L'} g^{aK'}_{J} g^{bI'}_{J'} M_{II'} M^{JJ'} 
   M_{KK'} M^{LL'}
   H_{\Fbar\Fbar VV\Fbar\Fbar}(I,J,a,b,K,L)
\nonumber \\ &&              
   +\frac{i}{3} g^{abc} g^{aI}_J g^{cK}_I g^{bJ}_K 
     H_{FFFVVV}(I,J,K,a,b,c)           
\nonumber \\ &&   
   + i g^{abc} g^{aI}_J g^{cK}_I g^{bK'}_{J'} M^{JJ'} M_{KK'} 
     H_{F\Fbar\Fbar VVV}(I,J,K,a,b,c) 
\nonumber \\ &&  
   + \frac{1}{2} g^{aK}_{I} g^{aJ}_{K} g^{bL}_{J} g^{bI}_{L}
   K_{FFFVVF}(I,J,K,a,b,L)                                                                                                                                                                                        
\nonumber \\ &&  
   + \frac{1}{2} g^{aK}_{I} g^{aK'}_{J} g^{bJ}_{L} g^{bI}_{L'} M_{KK'} M^{LL'}
   K_{FF\Fbar VV\Fbar}(I,J,K,a,b,L)                                                            
\nonumber \\ &&  
   + \frac{1}{2} g^{aK}_{I} g^{aJ}_{K} g^{bJ'}_{L} g^{bL}_{I'} M^{II'} M_{JJ'}
   K_{\Fbar\Fbar FVVF}(I,J,K,a,b,L)                                                            
\nonumber \\ &&  
   + \left (g^{aK}_{I} g^{aK'}_{J} g^{bJ}_{L} g^{bL}_{I'} M^{II'} M_{KK'}
   + \cc \right )
   K_{\Fbar F\Fbar VVF}(I,J,K,a,b,L)                                                            
\nonumber \\ &&  
   + \frac{1}{4} \left (g^{aK}_{I} g^{aK'}_{J} g^{bL}_{J'} g^{bL'}_{I'}
   M^{II'} M^{JJ'} M_{KK'} M_{LL'} + \cc \right )
   K_{\Fbar\Fbar\Fbar VV\Fbar}(I,J,K,a,b,L)                                                                     
\nonumber \\ &&  
   + \frac{1}{4} g^{aI}_{J} g^{bJ}_{I} g^{aK}_{L} g^{bL}_{K}
   K_{VVFFFF}(a,b,I,J,K,L)                            
\nonumber \\ &&  
   + \frac{1}{2} g^{aI}_{J} g^{bJ}_{I} g^{aK}_{L} g^{bK'}_{L'} M_{KK'} M^{LL'}
   K_{VVFF\Fbar\Fbar}(a,b,I,J,K,L)                            
\nonumber \\ &&  
   + \frac{1}{4} g^{aI}_{J} g^{bI'}_{J'} g^{aK}_{L} g^{bK'}_{L'} 
   M_{II'} M^{JJ'} M_{KK'} M^{LL'}
   K_{VV\Fbar\Fbar\Fbar\Fbar}(a,b,I,J,K,L)                            
\nonumber \\ &&  
   + \frac{1}{4} g^{acd} g^{bcd} g^{aI}_{J} g^{bJ}_{I}
   K_{\mbox{gauge},\>FF}(a,b,c,d,I,J)                        
\nonumber \\ &&  
   + \frac{1}{4} g^{acd} g^{bcd} g^{aI}_{J} g^{bI'}_{J'} M_{II'} M^{JJ'}
   K_{\mbox{gauge},\>\Fbar\Fbar}(a,b,c,d,I,J)  
.                      
\eeq
Again the loop integral functions appearing here are presented explicitly in the ancillary
electronic file {\tt functions.anc}.  

\subsection{Scalar, fermion, and vector contributions\label{subsec:V3SFV}}

The contributions that involve all three of scalars, fermions, and vectors are:
\beq
V^{(3)}_{SFV} &=& 
   \frac{1}{2} g^{aI}_J g^{aL}_{K} Y_{jIL} Y^{jJK} 
   H_{FFSVFF}(I,J,j,a,K,L)  
\nonumber \\ &&      
   + g^{aI}_J g^{aK}_{L} Y_{jIL'} Y^{jJK'} M_{KK'} M^{LL'} 
   H_{FFSV\Fbar\Fbar}(I,J,j,a,K,L)  
\nonumber \\ &&      
   + \frac{1}{2} \left (g^{aI}_J g^{aL}_{K} Y_{jIL} Y_{jJ'K'} M^{JJ'} M^{KK'} 
   + \cc \right )
   H_{F\Fbar SV\Fbar F}(I,J,j,a,K,L)  
\nonumber \\ &&      
   + \frac{1}{2} \left (g^{aI}_J g^{aK}_{L} Y_{jIL'} Y_{jJ'K} M^{JJ'} M^{LL'} 
   + \cc \right )
   H_{F\Fbar SVF\Fbar}(I,J,j,a,K,L)  
\nonumber \\ &&      
   + \frac{1}{2} 
   g^{aI}_J g^{aL}_{K} Y^{jI'L'} Y_{jJ'K'} M_{II'} M^{JJ'} 
   M^{KK'} M_{LL'} 
   H_{\Fbar\Fbar SV\Fbar\Fbar}(I,J,j,a,K,L)  
\nonumber \\ &&         
   + i g^{ajk} g^{aI}_J  Y_{kIK} Y^{jJK} 
   H_{FFFVSS}(I,J,K,a,j,k)   
\nonumber \\ &&  
   + \left (i g^{ajk} g^{aI}_J  Y_{kIK}  Y_{jJ'K'} M^{JJ'} M^{KK'}
   + \cc \right )
   H_{F\Fbar\Fbar VSS}(I,J,K,a,j,k)  
\nonumber \\ &&      
   +i g^{ajk} g^{aI}_J Y^{kI'K} Y_{jJ'K} M_{II'} M^{JJ'} 
   H_{\Fbar\Fbar FVSS}(I,J,K,a,j,k)   
\nonumber \\ &&  
   + \left (g^{abj} g^{aJ}_{K} g^{bK}_{I} Y^{jIJ'} M_{JJ'} + \cc \right )
   H_{F\Fbar FSVV}(I,J,K,j,a,b)   
\nonumber \\ &&  
   + \frac{1}{2} \left (g^{abj} g^{aK}_{J} g^{bK'}_{I} Y^{jIJ} M_{KK'} + \cc \right )
   H_{FF\Fbar SVV}(I,J,K,j,a,b)   
\nonumber \\ &&  
   + \frac{1}{2} \left (g^{abj} g^{aJ}_{K} g^{bI}_{K'} Y^{jI'J'} M_{II'} M_{JJ'} M^{KK'} 
   + \cc \right )
   H_{\Fbar\Fbar\Fbar SVV}(I,J,K,j,a,b)   
\nonumber \\ &&  
   + g^{aL}_{I} g^{aJ}_{L} Y^{jIK} Y_{jJK} 
   K_{FFFSVF}(I,J,K,j,a,L) 
\nonumber \\ &&  
   + \frac{1}{2} \left(g^{aL}_{I} g^{aL'}_{J} Y^{jIK} Y^{jJK'} M_{KK'} M_{LL'} 
   + \cc \right ) 
   K_{FF\Fbar SV\Fbar}(I,J,K,j,a,L) 
\nonumber \\ &&  
   + g^{aI}_{L} g^{aL}_{J} Y^{jI'K} Y_{jJ'K} M_{II'} M^{JJ'} 
   K_{\Fbar\Fbar FSVF}(I,J,K,j,a,L) 
\nonumber \\ &&  
   + \left (g^{aI}_{L} g^{aL}_{J} Y^{jI'K} Y^{jJK'} M_{II'} M_{KK'} +\cc\right )
   K_{\Fbar F\Fbar SVF}(I,J,K,j,a,L) 
\nonumber \\ &&  
   + \left (g^{aI}_{L} g^{aJ}_{L'} Y^{jI'K} Y_{jJK} M_{II'} M^{LL'} +\cc\right )
   K_{\Fbar FF SV\Fbar}(I,J,K,j,a,L) 
\nonumber \\ &&  
   + \frac{1}{2} \left (g^{aI}_{L} g^{aJ}_{L'} Y^{jI'K} Y^{jJ'K'} M_{II'} M_{JJ'}
   M_{KK'} M^{LL'} +\cc\right )
   K_{\Fbar\Fbar\Fbar SV\Fbar}(I,J,K,j,a,L) 
\nonumber \\ &&  
   + \frac{1}{2} g^{ajl} g^{alk} Y^{jIJ} Y_{kIJ}
   K_{SSSVFF}(j,k,l,a,I,J) 
\nonumber \\ &&  
   + \frac{1}{4} \left ( g^{ajl} g^{alk} Y^{jIJ} Y^{kI'J'} M_{II'} M_{JJ'} + \cc \right )
   K_{SSSV\Fbar\Fbar}(j,k,l,a,I,J) 
\nonumber \\ &&  
   + \frac{1}{4} g^{abj} g^{abk} Y^{jIJ} Y_{kIJ}
   K_{SSVVFF}(j,k,a,b,I,J) 
\nonumber \\ &&  
   + \frac{1}{8} \left (g^{abj} g^{abk} Y^{jIJ} Y^{kI'J'} M_{II'} M_{JJ'} + \cc \right )
   K_{SSVV\Fbar\Fbar}(j,k,a,b,I,J) 
\nonumber \\ &&  
   + \frac{1}{4} g^{ajk} g^{bkj} g^{aI}_{J} g^{bJ}_{I}
   K_{VVSSFF}(a,b,j,k,I,J) 
\nonumber \\ &&  
   + \frac{1}{4} g^{ajk} g^{bkj} g^{aI}_{J} g^{bI'}_{J'} M_{II'} M^{JJ'}
   K_{VVSS\Fbar\Fbar}(a,b,j,k,I,J) 
\nonumber \\ &&  
   + \frac{1}{2} g^{acj} g^{bcj} g^{aI}_{J} g^{bJ}_{I}
   K_{VVSVFF}(a,b,j,c,I,J) 
\nonumber \\ &&  
   + \frac{1}{2} g^{acj} g^{bcj} g^{aI}_{J} g^{bI'}_{J'} M_{II'} M^{JJ'}
   K_{VVSV\Fbar\Fbar}(a,b,j,c,I,J) 
.
\eeq
As before, 
the loop integral functions appearing here are presented explicitly in the ancillary
electronic file {\tt functions.anc}.  

\subsection{Pure vector and ghost contributions\label{subsec:V3V}}

Finally, the contributions that involve only vector bosons and ghost fields 
can be written in terms of a couple of loop integral functions:
\beq
V^{(3)}_{V} &=&   
   \frac{1}{24} g^{abd} g^{bce} g^{acf} g^{def}
     H_{\mbox{gauge}}(a,b,c,d,e,f) 
\nonumber \\ &&        
   + \frac{1}{16} g^{acd} g^{bcd} g^{aef} g^{bef}
     K_{\mbox{gauge}}(a,b,c,d,e,f) 
.
\label{eq:V3loopVgeneral}
\eeq
These contributions vanish except when the gauge symmetry associated
with the vectors is non-Abelian and (at least partly) spontaneously broken
by the scalar background fields. 
The loop integral functions appearing here are again presented explicitly in the ancillary
electronic file {\tt functions.anc}.

\subsection{Comments on the general results\label{subsec:comments}}

Equations (\ref{eq:V3loopSgeneral})-(\ref{eq:V3loopVgeneral}) 
constitute the complete \MSbar three-loop effective potential
contributions for a generic renormalizable quantum field theory with Landau gauge fixing.
However, for the specialization to any particular theory with massless gauge bosons, 
there is still a little processing to do in order to obtain 
the effective potential in practice. This is because each loop integral function 
involving a vector field with squared mass $x$ will contain a factor of $1/x$, 
which naively might appear to be have a pole singularity in the massless limit. 
This is due to the structure of the Landau gauge vector 
propagator proportional to 
\beq
\frac{\eta^{\mu\nu}}{p^2 + x} - \frac{p^\mu p^\nu}{p^2(p^2 + x)},
\eeq
(using a metric of signature $-$,$+$,$+$,$+$) where the second term has a partial fraction
decomposition proportional to 
\beq
\frac{1}{x} \left (\frac{1}{p^2 + x} - \frac{1}{p^2}\right ).
\eeq
Massless gauge bosons, with their potential infrared problems, 
are treated here by putting
$x = \delta$ and taking the limit $\delta \rightarrow 0$.
The factors of $1/\delta$ actually always cancel in the limit,
leaving behind either a finite result or singularities in each diagram that are at most 
logarithmic in $\delta$. However, demonstrating this starting from the general
loop integral functions appearing in the ancillary file {\tt functions.anc}, 
and finding the limits, requires using the expansions of the 
basis integrals in small squared masses, as given in the 
ancillary file {\tt expzero.anc}. This can be performed systematically
on a case-by-case basis, as will be done below for the example of the Standard Model. 
While the pole singularities in $\delta$ always cancel at the level of the loop integral functions, 
logarithmic singularities as $\delta \rightarrow 0$ 
can occur, but only when there is a doubled propagator, which means
the diagram topology is $K$, $J$, or $L$ (see Figure \ref{fig:diagramtopologies}) 
with the first two squared mass arguments 
both equal to $\delta$. The $\lnbar(\delta)$ singularities can then be obtained 
with the help of expansion formulas of the type
in the ancillary file {\tt expzero.anc}.
Cancellations of the infrared singularities associated with massless vector bosons
in the full effective potential occurs after 
summing the contributions of distinct diagrams, 
as will be illustrated below for the Standard Model.

Note also 
that the function $K_{\Fbar\Fbar FSSF}(x,w,u,z,y,v)$ in eq.~(\ref{eq:KffFSSF}) 
contains a factor $1/wx$, which 
naively might appear to be singular when either $w$ or $x$ approaches 0. However,
this is illusory; 
$K_{\Fbar\Fbar FSSF}(\delta,w,u,z,y,v)$ and $K_{\Fbar\Fbar FSSF}(x,\delta,u,z,y,v)$
and $K_{\Fbar\Fbar FSSF}(\delta,\delta,u,z,y,v)$ 
are each finite as $\delta \rightarrow 0$, as one can check by using 
the expansions given in the ancillary
file {\tt expzero.anc}. Furthermore, this
function appears in $V^{(3)}$ 
multiplied by $\sqrt{wx}$ [because of the fermion mass insertions
multiplying it in eq.~(\ref{eq:V3loopSFgeneral})]. 
Therefore, it does not contribute at all when $w$ and/or $x$ is zero. More 
generally, the contribution from every integral function 
with an $\Fbar$ subscript listed in eq.~(\ref{eq:genloopfunctions}) vanishes
when the corresponding fermion squared mass is taken to 0. Also,
$K_{FFFSSF}(\delta,\delta,u,z,y,v)$ and 
$K_{FF\Fbar SS\Fbar}(\delta,\delta,u,z,y,v)$, etc., have no 
$\lnbar(\delta)$ singularities. 
There are no infrared problems associated with massless fermions.

For checking purposes, it is useful to be able to take derivatives of the loop integral
functions with respect to the \MSbar renormalization scale $Q$. First, for the
basis functions and related functions, one has from ref.~\cite{Martin:2016bgz}:
\beq
Q \frac{\partial}{\partial Q} A(x) &=& -2 x
,
\label{eq:QdQA}
\\
Q \frac{\partial}{\partial Q} \Abar(x,y) &=& 2
,
\label{eq:QdQAbar}
\\
Q \frac{\partial}{\partial Q} I(x,y,z) &=& 
2 [A(x) + A(y) + A(z) -x-y-z]
,
\label{eq:QdQI}
\\
Q \frac{\partial}{\partial Q} \Ibar(w,x,y,z) &=& 2 + 2 \Abar(w,x)
,
\label{eq:QdQIbar}
\\
Q \frac{\partial}{\partial Q} E(w,x,y,z) &=& 
2 [A(w) A(x) 
+ A(w) A(y) 
+ A(w) A(z) 
+ A(x) A(y) 
\nonumber \\ &&
+ A(x) A(z) + A(y) A(z)
+ w x + w y + w z + x y + x z + y z]
\nonumber \\ &&
+ (w-2x-2y-2z) A(w) 
+ (x-2w-2y-2z) A(x)
\nonumber \\ &&
+ (y-2w-2x-2z) A(y)
+ (z-2w-2x-2y) A(z)
\nonumber \\ &&
-9 (w^2 + x^2 + y^2 + z^2)/4
,
\label{eq:QdQE}
\\
Q \frac{\partial}{\partial Q} F(w,x,y,z) &=& 
2 [-A(x)-A(y)-A(z) +x+y+z-w] A(w)/w 
\nonumber \\ &&
+ 7w/2
,
\label{eq:QdQF}
\\
Q \frac{\partial}{\partial Q} \Fbar(w,x,y,z) &=&
2 [A(x) + A(y) + A(z) - A(w) -x-y-z -I(x,y,z)] 
\nonumber \\ &&
+ 7w/2 
,
\phantom{xxx}
\label{eq:QdQFbar}
\\
Q \frac{\partial}{\partial Q} G(v,w,x,y,z) &=& 
2 [I(v,w,x) + I(v,y,z) + A(w) + A(x) + A(y) + A(z) + v]
\nonumber \\ &&
 -4(w+x+y+z)
,
\label{eq:QdQG}
\\
Q \frac{\partial}{\partial Q} H(u,v,w,x,y,z) &=& 12 \zeta_3
,
\label{eq:QdQH}
\\
Q \frac{\partial}{\partial Q} K(u,v,w,x,y,z) &=& 
2 [\Ibar(u,v,w,x) + \Ibar(u,v,y,z) -1]
,
\label{eq:QdQK}
\eeq
These results can now be applied to obtain the $Q$ derivatives of the 89 
integral functions of eq.~(\ref{eq:genloopfunctions}). Again the results 
are rather lengthy, and so are consigned to an ancillary electronic file 
{\tt QdQ.anc} provided with this paper.

\section{The Wess-Zumino model\label{sec:WZ}}
\setcounter{equation}{0}
\setcounter{figure}{0}
\setcounter{table}{0}
\setcounter{footnote}{1}

In this section, we consider as an example (and a confidence-building consistency check) 
the supersymmetric Wess-Zumino model \cite{Wess:1974tw,Wess:1973kz}, with superpotential 
(for a review, see \cite{Martin:1997ns}):
\beq
W = \frac{m}{2} \Phi^2 + \frac{y}{6} \Phi^3,
\eeq
with real mass and coupling parameters $m$ and $y$. 
The chiral superfield $\Phi$ contains a 2-component 
fermion $\psi$ and a complex scalar field that one can write as
\beq
\phi + (R + i I)/\sqrt{2},
\eeq
where $\phi$ is a constant background field and $R,I$ are canonically normalized
real scalar fields. In the following, depending on context, 
the names of the component quantum fields will also be used as synonyms for their field-dependent squared masses:
\beq
R &=& m^2 + 3 y m \phi + 3 y^2 \phi^2/2,
\\
I &=& m^2 + y m \phi + y^2 \phi^2/2,
\\
\psi &=& (m + y \phi)^2.
\eeq
The non-vanishing
interaction couplings of the fields $R,I,\psi$ are given by
\beq
\lambda^{RRRR} = \lambda^{IIII} = 3 \lambda^{RRII} &=& 3 y^2/2,
\\
\lambda^{RRR} = 3 \lambda^{RII} &=& 3 y (m + y \phi)/\sqrt{2},\phantom{xxxxxxx}
\\
Y^{R\psi\psi} &=& y/\sqrt{2},
\\
Y^{I\psi\psi} &=& iy/\sqrt{2},
\eeq
and permutations 
$\lambda^{RIRI} = \lambda^{RIIR} = \lambda^{IRRI} = \lambda^{IRIR} = \lambda^{IIRR} =\lambda^{RRII}$, and 
$\lambda^{IRI} = \lambda^{IIR} = \lambda^{RII}$.
There are no vector fields in the Wess-Zumino model.

The tree-level potential for the background field $\phi$ is  
\beq
V^{(0)} &=& \phi^2 (m + y \phi/2)^2 .
\label{eq:WZVtree} 
\eeq
Plugging into the results of
eqs.~(\ref{eq:V1loopgeneral}), (\ref{eq:V2loopgeneral}), 
(\ref{eq:V3loopSgeneral}), and 
(\ref{eq:V3loopSFgeneral})
above gives the effective potential contributions at 
one, two, and three-loop orders for the Wess-Zumino model:
\beq
V^{(1)} &=& f(R) + f(I) - 2 f(\psi)
,
\label{eq:WZVoneloop}
\\
V^{(2)} &=& y^2 \Bigl [ 3 f_{SS}(R,R)/16 + 3 f_{SS}(I,I)/16 + f_{SS}(I,R)/8
+ 3 \psi f_{SSS}(R,R,R)/8 
\nonumber \\ &&
+ \psi f_{SSS}(I,I,R)/8
+ f_{FFS}(\psi,\psi,R)/4 
+ f_{FFS}(\psi,\psi,I)/4 
\nonumber \\ &&
+ \psi f_{\Fbar\Fbar S}(\psi,\psi,R)/4 
- \psi f_{\Fbar\Fbar S}(\psi,\psi,I)/4
\Bigr ]
,
\label{eq:WZVtwoloop}
\\
V^{(3)} &=& y^4 \psi^2 \Bigl [
27 H_{RRRRRR}/32 
+ H_{IIRRII}/32 
+ H_{IIIRRR}/8
+ 81 K_{RRRRRR}/64  
\nonumber \\ &&
+ 9 K_{RRIIRR}/32 
+ K_{RRIIII}/64   
+ K_{IIIRIR}/16 
+ H_{\psibar\psibar RR \psibar\psibar}/16
\nonumber \\ &&
- H_{\psibar\psibar IR \psibar\psibar}/8
+ H_{\psibar\psibar II \psibar\psibar}/16
+ H_{\psibar\psibar\psibar RRR}/4
- H_{\psibar\psibar\psibar IIR}/4 
\nonumber \\ &&
+ 9 K_{RRRR\psibar\psibar}/16
+ K_{RRII\psibar\psibar}/16
- K_{IIIR\psibar\psibar}/8
+ K_{RR\psibar\psibar\psibar\psibar}/16 
\nonumber \\ &&
+ K_{II\psibar\psibar\psibar\psibar}/16 
+ K_{\psibar\psibar\psibar RR \psibar}/8
- K_{\psibar\psibar\psibar IR \psibar}/4
+ K_{\psibar\psibar\psibar II \psibar}/8
\Bigr ]
\nonumber \\ &&
+ y^4 \psi \Bigl [ 
27 G_{RRRRR}/32
+ 3 G_{RIIRR}/16
+ G_{IIRIR}/8
+ 3 G_{RIIII}/32
\nonumber \\ &&
+ 27 J_{RRRRR}/32
+ 9 J_{RRRRI}/32
+ 3 J_{RRIIR}/32
+ J_{RRIII}/32
+ 3 J_{IIIRI}/16
\nonumber \\ &&
+ J_{IIIRR}/16
+ H_{\psi\psi RR\psibar\psibar}/4
+ H_{\psi\psi RI\psibar\psibar}/4
- H_{\psi\psi IR\psibar\psibar}/4
- H_{\psi\psi II\psibar\psibar}/4
\nonumber \\ &&
+ H_{\psi\psibar RR\psi\psibar}/8
+ H_{\psi\psibar IR\psi\psibar}/4
+ H_{\psi\psibar II\psi\psibar}/8
+ 3 H_{\psi\psi\psibar RRR}/4
- H_{\psi\psi\psibar RII}/4
\nonumber \\ &&
+ H_{\psi\psi\psibar IIR}/2
+ 9 K_{RRRR\psi\psi}/16
+ K_{RRII\psi\psi}/16
+ K_{IIIR\psi\psi}/8
\nonumber \\ &&
+ K_{RR\psi\psi\psibar\psibar}/8 
- K_{II\psi\psi\psibar\psibar}/8 
+ K_{\psi\psi\psibar RR \psibar}/8
- K_{\psi\psi\psibar IR \psibar}/4
\nonumber \\ &&
+ K_{\psi\psi\psibar II \psibar}/8
+ K_{\psibar\psibar\psi RR\psi}/8
+ K_{\psibar\psibar\psi IR\psi}/4
+ K_{\psibar\psibar\psi II\psi}/8
\nonumber \\ &&
+ K_{\psibar\psi\psibar RR\psi}/2
+ K_{\psibar\psi\psibar RI\psi}/2
- K_{\psibar\psi\psibar IR\psi}/2
- K_{\psibar\psi\psibar II\psi}/2
\nonumber \\ &&
+ 3 J_{RR\psibar\psibar R}/16
+ J_{RR\psibar\psibar I}/16
- J_{II\psibar\psibar R}/16
- 3 J_{II\psibar\psibar I}/16
\Bigr ]
\nonumber \\ &&
+ y^4 \Bigl [
H_{\psi\psi RR\psi\psi}/16
- H_{\psi\psi IR\psi\psi}/8
+ H_{\psi\psi II\psi\psi}/16
+ K_{RR\psi\psi\psi\psi}/16 
\nonumber \\ &&
+ K_{II\psi\psi\psi\psi}/16 
+ K_{\psi\psi\psi RR \psi}/8
+ K_{\psi\psi\psi IR \psi}/4
+ K_{\psi\psi\psi II \psi}/8
\nonumber \\ &&
+ 9 L_{RRRR}/64
+ 9 L_{IIII}/64 
+ 3 L_{IIIR}/32
+ L_{IIRR}/64
+ L_{RRII}/64
\nonumber \\ &&
+ 3 L_{RRIR}/32
+ 3 E_{RRRR}/64 + 3 E_{IIII}/64 + E_{IIRR}/32
\nonumber \\ &&
+ 3 J_{RR\psi\psi R}/16
+ J_{RR\psi\psi I}/16
+ J_{II\psi\psi R}/16
+ 3 J_{II\psi\psi I}/16
\Bigr ]
.
\label{eq:WZVthreeloop}
\eeq
In the latter equation, I have used a short-hand notation, such that, for example,
$H_{IIRRII} \equiv H_{SSSSSS}(I,I,R,R,I,I)$ and
$K_{\psibar\psi\psibar RI\psi} \equiv K_{\Fbar F\Fbar SSF}(\psi,\psi,\psi,R,I,\psi)$.

As a non-trivial consistency 
check, consider the renormalization group scale invariance condition for the
effective potential, as expressed by eq.~(\ref{eq:QdQVexpanded}), with $X = y,m,\phi$.
From\footnote{Some other references
had given incorrect results for the 3-loop beta functions of the Wess-Zumino model.}
refs.~\cite{Avdeev:1982jx,Jack:1996qq},
\beq
{\beta^{(1)}_y}/{3y} \>=\> {\beta^{(1)}_m}/{2 m} \>=\> 
-{\beta^{(1)}_\phi}/{\phi} &=& y^2/2
,
\label{eq:betasy}
\\
{\beta^{(2)}_y}/{3y} \>=\> {\beta^{(2)}_m}/{2 m} \>=\> 
-{\beta^{(2)}_\phi}/{\phi} &=& -y^4/2
,
\\
{\beta^{(3)}_y}/{3y} \>=\> {\beta^{(3)}_m}/{2 m} \>=\> 
-{\beta^{(3)}_\phi}/{\phi} &=& (3\zeta_3/2 + 5/8) y^6
.
\label{eq:betasphi}
\eeq
Using eqs.~(\ref{eq:betasy})-(\ref{eq:betasphi}),
one finds from eq.~(\ref{eq:WZVtree}):
\beq
\sum_X \beta_X^{(1)} \frac{\partial}{\partial X} V^{(0)} &=&  
y^2 \phi^2 (m + y\phi/2)^2,
\label{eq:WZbeta1V0}
\\
\sum_X \beta_X^{(2)} \frac{\partial}{\partial X} V^{(0)} &=&  
-y^4 \phi^2 (m + y\phi/2)^2,
\\
\sum_X \beta_X^{(3)} \frac{\partial}{\partial X} V^{(0)} &=&  
(3 \zeta_3 + 5/4) y^6 \phi^2 (m + y\phi/2)^2,
\eeq
and from eq.~(\ref{eq:WZVoneloop}), also using eqs.~(\ref{eq:dA}):
\beq
\sum_X \beta_X^{(1)} \frac{\partial}{\partial X} V^{(1)} &=&  
y^2 (m^2 + 3 y m \phi + 3 y^2 \phi^2/2) A(R)
+ y^2 (m^2 + y m \phi + y^2 \phi^2/2) A(I)
\nonumber \\ && 
-2 y^2 (m + y\phi)^2 A(\psi) ,
\\
\sum_X \beta_X^{(2)} \frac{\partial}{\partial X} V^{(1)} &=&  
-y^4 (m^2 + 3 y m \phi + 3 y^2 \phi^2/2) A(R)
-y^4 (m^2 + y m \phi + y^2 \phi^2/2) A(I)
\phantom{xxxx}
\nonumber \\ && 
+ 2 y^4 (m + y\phi)^2 A(\psi) ,
\eeq
and from eq.~(\ref{eq:WZVtwoloop}), also using eqs.~(\ref{eq:dA}) and (\ref{eq:dI}):
\beq
\sum_X \beta_X^{(1)} \frac{\partial}{\partial X} V^{(2)} &=&  
\frac{7 y^4}{8} \Bigl [
-3 (m + y\phi)^2 I(R,R,R) 
+ (6 m^2 + 10 y m \phi + 5 y^2 \phi^2) I(\psi,\psi,R)
\nonumber \\ && 
- (m + y\phi)^2  I(R,I,I)
- (2 m^2 + 2 y m \phi + y^2 \phi^2) I(\psi,\psi,I)
+ 3 A(R)^2/2 
\phantom{xxxx}
\nonumber \\ && 
+ 3 A(I)^2/2 
+ A(R) A(I) 
-4 A(R) A(\psi) 
-4 A(I) A(\psi) 
+ 4 A(\psi)^2 
\Bigr ]
\nonumber \\ && 
+ y^4 (m^2 + 3 y m\phi + 3 y^2 \phi^2/2) A(R)
+ y^4 (m^2 + y m\phi + y^2 \phi^2/2) A(I)
\nonumber \\ && 
-2 y^4 (m + y \phi)^2 A(\psi)
- y^6 \phi^2 (m + y \phi/2)^2
.
\eeq
Meanwhile, from eqs.~(\ref{eq:WZVoneloop}), (\ref{eq:WZVtwoloop}), and (\ref{eq:WZVthreeloop}),
using eqs.~(\ref{eq:QdQA})-(\ref{eq:QdQK}), one obtains:
\beq
Q \frac{\partial}{\partial Q} V^{(1)} &=& -y^2 \phi^2 (m + y\phi/2)^2,
\\
Q \frac{\partial}{\partial Q} V^{(2)} &=&
- y^2 (m^2 + y m \phi + y^2\phi^2/2) A(I)
- y^2 (m^2 + 3 y m \phi + 3 y^2\phi^2/2) A(R) 
\nonumber \\ && 
+ 2 y^2 (m + y\phi)^2  A(\psi) 
+ y^4 \phi^2 (m + y \phi/2)^2 
,
\\
Q \frac{\partial}{\partial Q} V^{(3)} &=& \frac{7y^4}{8} \Bigl [
(2 m^2 + 2 y m \phi + y^2 \phi^2) I(\psi,\psi,I)
- (6 m^2 + 10 y m \phi + 5 y^2 \phi^2) I(\psi,\psi,R)
\nonumber \\ && 
+ 3 (m + y\phi)^2 I(R,R,R) + (m + y\phi)^2  I(R,I,I) 
- 3 A(R)^2/2 
- 3 A(I)^2/2
\nonumber \\ && 
-A(R) A(I) - 4 A(\psi)^2
+4  A(\psi) A(R)  +4  A(\psi) A(I) \Bigr ]
\nonumber \\ && 
- y^6 \phi^2 (m + y\phi/2)^2 (3 \zeta_3 + 1/4)
.
\label{eq:WZQdQV3}
\eeq
Now eqs.~(\ref{eq:WZbeta1V0})-(\ref{eq:WZQdQV3}) can be plugged in to verify
that eq.~(\ref{eq:QdQVexpanded}) indeed holds for each of $\ell=1,2,3$.

As another check, recall that at a supersymmetric minimum of the tree-level potential, 
the full effective potential must vanish at each order in perturbation theory
\cite{Zumino:1974bg}. 
There are two supersymmetric minima of $V^{(0)}$, namely 
$\phi = 0$ and $\phi = -2m/y$. At each of these, one has equality of the field-dependent
squared masses: $x \equiv R = I = \psi = m^2$. 
It is now straightforward to plug this into equations 
(\ref{eq:WZVoneloop}), 
(\ref{eq:WZVtwoloop}), and
(\ref{eq:WZVthreeloop}), to verify that each of $V^{(1)}$, $V^{(2)}$, and $V^{(3)}$ also vanishes at the supersymmetric minima. This relies on non-trivial
cancellations between the different loop integral functions defined in eqs.~(\ref{eq:defHSSSSSS})-(\ref{eq:defESSSS}) and 
(\ref{eq:defHFFfSSS})-(\ref{eq:defJSSffS}), which 
become apparent upon putting everything 
in terms of the basis integral functions $H(x,x,x,x,x,x)$, $K(x,x,x,x,x,x)$,
$G(x,x,x,x,x)$, $F(x,x,x,x)$, $I(x,x,x)$, $\Ibar(x,x,x,x)$, $A(x)$, and $\Abar(x,x)$.

\section{The Standard Model\label{sec:SM}}
\setcounter{equation}{0}
\setcounter{figure}{0}
\setcounter{table}{0}
\setcounter{footnote}{1}

\subsection{Standard Model effective potential at three-loop order \label{subsec:SMeffpot}}

In this section, I will consider the complete
three-loop effective potential for the Standard Model as another application of
the general results. The full 
two-loop effective potential for the Standard Model was found in ref.~\cite{Ford:1992pn}.
The leading three-loop parts,
in the limit that the QCD coupling and the top-quark Yukawa coupling are large
compared to all other couplings, were found in ref.~\cite{Martin:2013gka}. 
The four-loop contribution at leading order in QCD
is also known \cite{Martin:2015eia}.

The tree-level potential for the Standard Model is given by
\beq
V &=& \Lambda + m^2 \Phi^\dagger \Phi + \lambda (\Phi^\dagger \Phi)^2,
\eeq
where $\Lambda$ is a field-independent constant energy density necessary for
renormalization scale invariance, $m^2$ is a negative Higgs squared mass parameter,
and $\lambda$ is the Higgs self-coupling. The Higgs complex doublet scalar field is
\beq
\Phi = 
\begin{pmatrix}
[\phi + H + i G^0]/\sqrt{2}
\\
G^+
\end{pmatrix}
,
\eeq
where $\phi$ is the constant background field, $H$ is the Higgs boson field,
and $G^0$ and $G^\pm$ are Goldstone bosons.
The $\phi$-dependent squared masses 
of the Higgs boson and the Goldstone bosons (both neutral and charged) are
\beq
H &=& m^2 + 3 \lambda \phi^2,
\\
G &=& m^2 + \lambda \phi^2,
\eeq
and the other non-zero squared masses are
\beq
t &=& y_t^2 \phi^2/2,
\\
W &=& g^2 \phi^2/4,
\\
Z &=& (g^2 + g^{\prime 2}) \phi^2/4.
\eeq
where $y_t$ is the top-quark Yukawa coupling and $g,g'$ are the electroweak gauge couplings. The Yukawa couplings of the bottom quark and other fermions
are quite negligible, and are therefore taken to vanish.

The field content of the electroweak Standard Model with $n_G$ generations, compatible
with the conventions of section \ref{sec:setup}, is:
\beq
\mbox{Real scalars:}&\>\>\>\>&H, G^0, G_R, G_I
\\
\mbox{2-component fermions:}
&\>\>\>\>&t,\,\bar t,\, b,\, \bar b,\, \tau,\, \bar \tau,\, \nu_\tau
\>+\> (n_G - 1) \times \left ( 
u,\, \bar u,\, d,\, \bar d,\, e,\, \bar e,\, \nu_e \right ),\phantom{xxxx}
\\
\mbox{Real vectors:}&\>\>\>\>&\gamma, Z, W_R, W_I
\eeq
Here we have written the complex Goldstone scalar bosons and $W$ vector bosons in terms
of real components as $G^\pm = (G_R \pm i G_I)/\sqrt{2}$ and 
$W^\pm = (W_R \pm i W_I)/\sqrt{2}$. 
The unbarred fermion fields are $SU(2)_L$ doublets,
and the barred fermion fields are $SU(2)_L$ singlets. 
(Not shown explicitly are the color multiplicity for quarks, 
or the massless gluon vector fields.)

To facilitate an automated calculation of the 3-loop effective potential,
it is useful to have at hand a list of the non-vanishing field-dependent 
couplings of these mass eigenstate fields.  
There are scalar cubic interactions of the type $\lambda^{jkl}$:
\beq
\lambda^{HHH} &=& 6 \lambda \phi,
\label{eq:SMhhhcoupling}
\\
\lambda^{H G^0 G^0} &=& \lambda^{H G_R G_R} \>=\> \lambda^{H G_I G_I} \>=\> 2 \lambda \phi,
\eeq
and scalar quartic couplings of the type $\lambda^{jklm}$:
\beq
\lambda^{HHHH} &=& 
\lambda^{G^0 G^0 G^0 G^0} \>=\> 
\lambda^{G_R G_R G_R G_R} \>=\> 
\lambda^{G_I G_I G_I G_I} \>=\> 6 \lambda ,
\\
\lambda^{HH G^0 G^0} &=& 
\lambda^{HH G_R G_R} = 
\lambda^{HH G_I G_I} =
\lambda^{G^0 G^0 G_R G_R} = \lambda^{G^0 G^0 G_I G_I} =
\lambda^{G_R G_R G_I G_I} = 2 \lambda ,\phantom{xxx} 
\eeq
as well as permutations determined by the symmetry under interchange of any 
two scalars.
The non-vanishing Yukawa couplings of the type $Y^{jIJ}$ are given by
\beq
Y^{H t\bar t} = -Y^{G_R b\bar t} = 
-i Y^{G^0 t\bar t} = i Y^{G_I b \bar t} &=& y_t/\sqrt{2},
\eeq
which are symmetric under interchange of the last two (fermionic) indices. 
(All Yukawa couplings other than for the top-quark are neglected.)
The interactions of the electroweak vector bosons with the quarks and leptons 
are given by couplings of the type $g^{aJ}_I$:
\beq
g^{Z f}_f &=& (-I_f g^2 + Y_f g^{\prime 2})/\sqrt{g^2 + g^{\prime 2}}
,
\\
g^{Z \bar f}_{\bar f} &=& -Q_f g^{\prime 2}/\sqrt{g^2 + g^{\prime 2}}
,
\\
g^{\gamma f}_f &=& -g^{\gamma \bar f}_{\bar f} = -Q_f e
,
\eeq
where 
\beq
e  &=& g g'/\sqrt{g^2 + g^{\prime 2}}
,
\eeq
and
$Q_u = 2/3$ and $Q_d = -1/3$ and $Q_{\nu} = 0$ and $Q_e = -1$, and
$I_u = I_{\nu} = 1/2$ and $I_d = I_e = -1/2$, and $Y_f = Q_f - I_f$ for each $f$,
and
\beq
g^{W_R u}_d = g^{W_R d}_u = g^{W_R \nu}_e = g^{W_R e}_\nu = -g/2,
\\
g^{W_I d}_u = -g^{W_I u}_d = g^{W_I e}_\nu = -g^{W_I \nu}_e = -ig/2.
\eeq
There are also vector-vector-scalar couplings of the type $g^{abj}$,
\beq
g^{\gamma W_R G_R} &=& g^{\gamma W_I G_I} \>=\> 
g e \phi/2,
\\
g^{Z W_R G_R} &=& g^{Z W_I G_I} \>=\> 
-g' e \phi/2,
\\
g^{W_R W_R H} &=& g^{W_I W_I H} \>=\> g^2 \phi/2,
\\
g^{ZZH} &=& (g^2 + g^{\prime 2}) \phi/2 ,
\eeq
with symmetry under interchange of the first two (vector) indices.
The vector-scalar-scalar couplings of the type $g^{ajk}$ are
\beq
g^{\gamma G_R G_I} &=& e,
\\
g^{Z G^0 H} &=& \sqrt{g^2 + g^{\prime 2}}/2,
\\
g^{Z G_R G_I} &=& (g^2 - g^{\prime 2})/(2 \sqrt{g^2 + g^{\prime 2}}),
\\
g^{W_R G_R G^0} &=& g^{W_R H G_I} \>=\> g^{W_I G_I G^0} \>=\> g^{W_I G_R H} \>=\>  g/2,
\label{eq:SMVSScouplings}
\eeq
with others determined by the anti-symmetry with 
respect to interchange of the last two (scalar) indices. There are also 
vector-vector-scalar-scalar couplings of the type $g^{abjk}$,
determined in terms of these by eq.~(\ref{eq:VVSScouplings}).
Finally there are the totally anti-symmetric vector-vector-vector couplings defined by:
\beq
g^{\gamma W_R W_I} &=& e,
\\
g^{Z W_R W_I} &=& g^2/\sqrt{g^2 + g^{\prime 2}}.
\eeq

The tree-level and one-loop contributions to the effective potential are:
\beq
V^{(0)} &=& \Lambda + m^2 \phi^2/2 + \lambda \phi^4/4,
\\
V^{(1)} &=& 3 f(G) + f(H) - 12 f(t) + 6 f_V(W) + 3 f_V(Z),
\eeq
with functions $f(x)$ and $f_V(x)$ defined in eqs.~(\ref{eq:deff}) and (\ref{eq:deffV}).
The two-loop contribution is given by \cite{Ford:1992pn}:
\beq
V^{(2)} &=& 
\frac{3}{4} \lambda \left [f_{SS}(H,H) + 2 f_{SS}(G,H) + 5 f_{SS}(G,G) \right ]
\nonumber \\ &&
+ 3 \lambda^2 \phi^2 [f_{SSS}(H,H,H) 
+ f_{SSS}(G,G,H)]
+ \frac{3 y_t^2}{2} \Bigl [
f_{FFS}(t,t,H) + t f_{\Fbar\Fbar S}(t,t,H)
\nonumber \\ &&
+ f_{FFS}(t,t,G) - t f_{\Fbar\Fbar S}(t,t,G)
+ 2 f_{FFS}(0,t,G)
\Bigr ]
+ \frac{g^2 + \gptwo}{8} f_{VSS}(Z,G,H)
\nonumber \\ &&
+ \frac{(g^2 - \gptwo)^2}{8 (g^2 + \gptwo)} f_{VSS}(Z,G,G)
+ \frac{g^2}{4} \left [ f_{VSS}(W,G,H) + f_{VSS}(W,G,G) \right]
\nonumber \\ &&
+ \frac{e^2}{2} f_{VSS}(0,G,G)
+ \frac{g^4 \phi^2}{8} f_{VVS}(W,W,H)
+ \frac{(g^2 + \gptwo)^2 \phi^2}{16} f_{VVS}(Z,Z,H)
\nonumber \\ &&
+\frac{e^2 \phi^2}{4}\left [\gptwo f_{VVS}(W,Z,G)
+ g^2 f_{VVS}(0,W,G)\right ]
\nonumber \\ &&
-\left [4 g_3^2 + 4 e^2/3
            \right ] t f_{\overline{F}\overline{F}V}(t,t,0)
+ g^2 \left [ 3 f_{FFV}(0,t,W) + (4 n_G - 3) f_{FFV}(0,0,W) \right ]/2
\nonumber \\ &&
+ \bigl [
(9 g^4 - 6 g^2 \gptwo + 17 \gpfour) f_{FFV}(t,t,Z)
+  8 \gptwo (3 g^2 - \gptwo) t f_{\overline{F}\overline{F}V}(t,t,Z)
\nonumber \\ &&
+ ((24 n_G - 9) g^4 + 6 g^2 \gptwo + (40 n_G - 17) \gpfour) f_{FFV}(0,0,Z)
\bigr ]/24 (g^2 + \gptwo)
\nonumber \\ &&
+ \frac{e^2}{2} f_{\rm gauge}(W,W,0)
+ \frac{g^4}{2 (g^2 + \gptwo)} f_{\rm gauge}(W,W,Z)
,
\label{eq:V2SM}
\eeq
where $n_G=3$ is the number of quark and lepton generations.
This result for $V^{(2)}$ for the Standard Model is an example of the 
application of the general result in eq.~(\ref{eq:V2loopgeneral}),
using the couplings listed above.
It is written here in terms of 
the functions $f_{SS}(x,y)$, $f_{SSS}(x,y,z)$, $f_{FFS}(x,y,z)$, 
$f_{\Fbar\Fbar S}(x,y,z)$, $f_{VVS}(x,y,z)$, $f_{FFV}(x,y,z)$, $f_{\Fbar\Fbar V}(x,y,z)$,
and $f_{\rm gauge}(x,y,z)$ defined in ref.~\cite{Martin:2001vx}, and the function
$f_{VSS}(x,y,z)$ defined in 
eqs.~(\ref{eq:deffVSS})-(\ref{eq:deffVSSxyz}) of the present paper, replacing
the functions $f_{SSV}$ and $f_{VS}$ of ref.~\cite{Martin:2001vx}.

The three-loop effective potential contribution in the Standard Model can now be 
obtained by applying the couplings given above in 
eqs.~(\ref{eq:SMhhhcoupling})-(\ref{eq:SMVSScouplings}) to the general
forms of eqs.~(\ref{eq:defHSSSSSS})-(\ref{eq:V3loopVgeneral}). 
The resulting expression contains 536
integral functions of the 89 types in 
eq.~(\ref{eq:genloopfunctions})  with specific assignments
of squared mass arguments $H$, $G$, $t$, $W$, $Z$, 
and $\delta$ (used as an infrared regulator for the squared
masses of gluons, photons, and quarks and leptons other than the top quark). 
The 536 functions are given, expanded in $\delta$ to retain
the $\lnbar(\delta)$ terms, but dropping all terms of order $\delta$, in an ancillary electronic file distributed with this paper,
called {\tt SMV3functions.anc}. 
Of these 536 functions, the following 23 vanish identically in the 
limit $\delta \rightarrow 0$:
\beq
&&
H_{FFVVFF}(0,0,0,0,0,0),\>\>\>
K_{FFFVVF}(0,0,0,0,0,0),\>\>\>
K_{VVFFFF}(0,0,0,0,0,0),\>\>\>
\nonumber \\ &&
K_{\Fbar \Fbar FSVF}(t, t, t, G, 0, t),\>\>\>
K_{\Fbar\Fbar FSVF}(t, t, t, H, 0, t),\>\>\>
K_{\Fbar \Fbar FSVF}(t, t, 0, G, 0, t),\>\>\>
\nonumber \\ &&
K_{\Fbar F\Fbar SVF}(t, t, t, G, 0, t),\>\>\>
K_{\Fbar F\Fbar SVF}(t, t, t, H, 0, t),\>\>\>
K_{FFFSVF}(t, t, t, G, 0, t),\>\>\>
\nonumber \\ &&
K_{FFFSVF}(t, t, t, H, 0, t),\>\>\>
K_{FFFSVF}(t, t, 0, G, 0, t),\>\>\>
K_{FFFSVF}(0, 0, t, G, 0, 0),\>\>\>
\nonumber \\ &&
K_{\Fbar \Fbar FVVF}(t, t, t, Z, 0, t),\>\>\>
K_{\Fbar \Fbar FVVF}(t, t, t, 0, W, 0),\>\>\>
K_{\Fbar \Fbar FVVF}(t, t, t, 0, 0, t),\>\>\>
\nonumber \\ &&
K_{\Fbar F\Fbar VVF}(t, t, t, Z, 0, t),\>\>\>
K_{\Fbar F\Fbar VVF}(t, t, t, 0, 0, t),\>\>\>
K_{FFFVVF}(t, t, t, Z, 0, t),\>\>\>
\nonumber \\ &&
K_{FFFVVF}(t, t, t, 0, W, 0),\>\>\>
K_{FFFVVF}(t, t, t, 0, 0, t),\>\>\>
K_{FFFVVF}(0, 0, t, W, 0, 0),\>\>\>
\nonumber \\ &&
K_{FFFVVF}(0, 0, 0, W, 0, 0),\>\>\>
K_{FFFVVF}(0, 0, 0, Z, 0, 0).
\eeq
The coefficients of the non-vanishing 513 functions that remain, and thus the
expression for $V^{(3)}$, are
given in another ancillary electronic file called {\tt SMV3.anc}. These
coefficients are built out of couplings 
$g_3$, $g$, $g'$, $y_t$, $\lambda$, and the background Higgs field $\phi$. 
In the text below, I will discuss explicitly the parts of $V^{(3)}$ that 
are leading order in QCD,
and also the parts involving infrared logarithms $\lnbar(\delta)$, where $\delta$
is used for the gluon and photon squared masses. (There are no $\lnbar(\delta)$
infrared divergences due to massless fermions, as discussed in subsection
\ref{subsec:comments}.)

Consider the contribution proportional to $g_3^4$. It is:  
\beq
V^{(3)}_{g_3^4} &=& g_3^4 N_c C_F \Bigl \{
(C_F - C_G/2) \Bigl [ 
\frac{1}{2} H_{FFVVFF}(t,t,0,0,t,t) 
- 2 t H_{FFVV\Fbar\Fbar}(t,t,0,0,t,t) 
\nonumber \\ &&
+ t H_{F\Fbar VVF\Fbar}(t,t,0,0,t,t) + 
\frac{t^2}{2} H_{\Fbar\Fbar VV\Fbar\Fbar}(t,t,0,0,t,t) 
\Bigr ]
\nonumber \\ &&
+ C_F \left [
t K_{FF\Fbar VV\Fbar} (t,t,t,0,0,t) + t^2 K_{\Fbar\Fbar\Fbar VV\Fbar} (t,t,t,0,0,t)
\right ]
\nonumber \\ &&
+ C_G \Bigl [
\frac{1}{3} H_{FFFVVV}(t,t,t,0,0,0) - t H_{F\Fbar\Fbar VVV}(t,t,t,0,0,0)
\nonumber \\ &&
+ \frac{1}{2} K_{{\rm gauge},FF}(0,0,0,0,t,t)
- \frac{1}{2} t K_{{\rm gauge},\Fbar\Fbar}(0,0,0,0,t,t)
\Bigr]
\nonumber \\ &&
+ T_F \left [
K_{VVFFFF}(\delta,\delta,t,t,t,t) - 2 t K_{VVFF\Fbar\Fbar}(\delta,\delta,t,t,t,t)
+ t^2 K_{VV\Fbar\Fbar\Fbar\Fbar}(\delta,\delta,t,t,t,t)
\right ]
\nonumber \\ &&
+ 2 (2 n_G - 1) T_F \left [ 
K_{VVFFFF}(0,0,0,0,t,t) - t K_{VVFF\Fbar\Fbar}(0,0,0,0,t,t)
\right ]
\Bigr \} ,
\label{eq:V3SMg34}
\eeq
where $C_G = N_c = 3$ and $C_F = 4/3$ and $T_F = 1/2$ for QCD, and $n_G = 3$
for the Standard Model. 
In the following, I will leave $n_G$ arbitrary, to allow for
more informative comparisons. In writing eq.~(\ref{eq:V3SMg34}), I have taken
advantage of the fact that $K_{FFFVVF}(t,t,t,0,0,t)$ and
$K_{\Fbar F\Fbar VVF}(t,t,t,0,0,t)$ and $K_{\Fbar\Fbar FVVF}(t,t,t,0,0,t)$ happen to 
vanish, even though those functions do not vanish for general squared-mass arguments.
The remaining loop integral functions for the special
squared mass arguments appearing in eq.~(\ref{eq:V3SMg34}) are also quite simple. The ones
without infrared gluon divergences (therefore setting $\delta=0$) are:
\beq
H_{FFVVFF}(t,t,0,0,t,t) &=& (225/2 - 208 \zetathree) t^2 - 85 t A(t) +
     6 A(t)^2 - 10 t^2 H(0, t, t, t, 0, t) 
,     
\phantom{xxxxx}
\\
H_{FFVV\Fbar\Fbar}(t,t,0,0,t,t) &=& (140/3 - 80 \zetathree) t - 40 A(t) +
     12 A(t)^2/t - 6 t H(0, t, t, t, 0, t) 
,
\\
H_{F\Fbar VVF\Fbar}(t,t,0,0,t,t) &=& (32 \zetathree -16) t + 10 A(t) +
     6 A(t)^2/t + 6 t H(0, t, t, t, 0, t) 
,
\\
H_{\Fbar\Fbar VV\Fbar\Fbar}(t,t,0,0,t,t) &=& 
16/3 + 16 \zetathree - 8 A(t)/t -
     10 H(0, t, t, t, 0, t) 
,
\\
K_{FF\Fbar VV\Fbar}(t,t,t,0,0,t) &=& 146 t/3 - 60 A(t) 
+ 18 A(t)^2/t - 18 A(t)^3/t^2
,
\\
K_{\Fbar\Fbar\Fbar VV\Fbar}(t,t,t,0,0,t) &=& -38/3 + 48 \zetathree + 10 A(t)/t -
     12 A(t)^2/t^2 - 6 A(t)^3/t^3
,
\\
H_{FFFVVV}(t,t,t,0,0,0) &=& (24 \zetathree - 233/8) t^2 + 117 t A(t)/4 - 27 A(t)^2/2  
,
\\
H_{F\Fbar\Fbar VVV}(t,t,t,0,0,0) &=&
(48 \zetathree - 136/3) t + 45 A(t) - 27 A(t)^2/t + 3 A(t)^3/t^2 
,
\\     
K_{{\rm gauge},FF}(0,0,0,0,t,t) &=& 
-283 t^2/6 + 40 t A(t) - 13 A(t)^2
,
\\
K_{{\rm gauge},\Fbar\Fbar}(0,0,0,0,t,t) &=& \left [(-296 - 208 \zetathree) t +
     281 A(t) - 97 A(t)^2/t + 26 A(t)^3/t^2 \right ]/3
,
\\
K_{VVFFFF}(0,0,0,0,t,t) &=& 49 t^2/6 - 7 t A(t) + 2 A(t)^2
,
\\
K_{VVFF\Fbar\Fbar}(0,0,0,0,t,t) &=& \left[
(56 + 32 \zetathree) t - 52 A(t) + 20 A(t)^2/t - 4 A(t)^3/t^2 \right ]/3
,
\eeq
while the ones that do individually have gluon infrared divergences are:
\beq
K_{VVFFFF}(\delta,\delta,t,t,t,t) &=& -(5 + 56 \zetathree) t^2 - 62 t A(t) -
     8 A(t)^2 - 8 A(t)^3/t 
\nonumber \\ && 
+ 12 t^2 (1 + A(t)/t)^2 \lnbar(\delta)     
,
\\
K_{VVFF\Fbar\Fbar}(\delta,\delta,t,t,t,t) &=& \left [
-(8 + 280 \zetathree) t  - 340 A(t) -
52 A(t)^2/t - 28 A(t)^3/t^2 \right ]/3 \phantom{xxxx}
\nonumber \\ && 
+ 12 t (1 + A(t)/t)^2 \lnbar(\delta)          
,
\\
K_{VV\Fbar\Fbar\Fbar\Fbar}(\delta,\delta,t,t,t,t) &=&
-152/3 - 56 \zetathree - 96 A(t)/t - 36 A(t)^2/t^2 - 8 A(t)^3/t^3 
\nonumber \\ && 
+ 12 (1 + A(t)/t)^2 \lnbar(\delta)
.
\eeq
It is now clear that the $\lnbar(\delta)$ contributions successfully 
cancel when these results
are put into eq.~(\ref{eq:V3SMg34}). The result is:
\beq
V^{(3)}_{g_3^4} &=& g_3^4 t^2 \Bigl [
8131/9 
- 84 n_G 
+ (320 - 256 n_G/3) \zetathree
+ (248 n_G/3 - 2834/3) A(t)/t
\nonumber \\ &&
+ (428 - 112 n_G/3) A(t)^2/t^2 
+ (32 n_G/3 - 216) A(t)^3/t^3 
- 16 H(0,t,t,t,0,t)/3
\Bigr ]
\phantom{xxxxx}
\\
&=& g_3^4 t^2 \Bigl [
22429/9 
- 644 n_G/3 
- 512 {\rm Li}_4(1/2)/3 
+ 64 \ln^2(2) [\pi^2 - \ln^2(2)]/9 
\nonumber \\ &&
+ 176 \pi^4/135
+ (288 - 256 n_G/3) \zetathree 
+ (32 \zetathree + 568 n_G/3 - 7346/3) \lnbar(t)
\nonumber \\ &&
+ (1076 - 208 n_G/3) \lnbar^2(t)
+ (32 n_G/3 - 216) \lnbar^3(t)
\Bigr ] ,
\eeq
where the analytical results for $A(t)$ and $H(0,t,t,t,0,t)$ 
have been used to obtain the last expression.
This agrees with the result found (using different methods, and in particular with
dimensional regularization of infrared divergences) in ref.~\cite{Martin:2013gka}.

Having demonstrated that the infrared divergences associated with doubled 
massless gluon propagators
cancel, let us now consider those coming from the massless photon.
First, from the results given in the ancillary electronic file 
{\tt SMV3.anc} and {\tt SMV3functions.anc}, one finds that 
$V^{(3)}$ contains QED contributions exactly analogous to the QCD ones
mentioned above:
\beq
\frac{16 e^4}{9} [
K_{VVFFFF}(\delta,\delta,t,t,t,t) 
-2 t K_{VVFF\Fbar\Fbar}(\delta,\delta,t,t,t,t)
+ t^2 K_{VV\Fbar\Fbar\Fbar\Fbar}(\delta,\delta,t,t,t,t)
],
\phantom{xx}
\eeq
where $\delta$ is now the infrared regulator squared mass of the photon.
The $\lnbar(\delta)$ parts of this cancel in the same way as in QCD. 

There are also contributions (given in {\tt SMV3.anc}) to $V^{(3)}$ from diagrams
involving the top quark, the $W$ boson, the charged Goldstone bosons, 
and doubled photon propagators, 
which individually behave like $\lnbar(\delta)$ as $\delta \rightarrow 0$. They can be 
grouped as:
\beq
\frac{4e^4}{3} [K_{{\rm gauge},FF}(\delta,\delta,W,W,t,t) 
- t K_{{\rm gauge},\Fbar\Fbar}(\delta,\delta,W,W,t,t) ] ,
\label{eq:lndeltatWA}
\eeq
and
\beq
\frac{2 e^4 g^2 \phi^2}{3} [
K_{VVSVFF}(\delta, \delta, G, W, t, t) - t K_{VVSV\Fbar\Fbar}(\delta, \delta, G, W, t, t)
]
.
\label{eq:lndeltatGWA}
\eeq
The $\lnbar(\delta)$ infrared divergent parts of these 
contributions can be extracted from the results given in the ancillary
electronic file {\tt SMV3functions.anc}:
\beq
K_{{\rm gauge},FF}(\delta,\delta,W,W,t,t) &\sim& 
\frac{3}{2}[W + 6 A(W)][t + A(t)]\, \lnbar(\delta),
\\
K_{{\rm gauge},\Fbar\Fbar}(\delta,\delta,W,W,t,t) &\sim& 
\frac{3}{2t}[W + 6 A(W)][t + A(t)]\, \lnbar(\delta),
\\
K_{VVSVFF}(\delta, \delta, G, W, t, t) &\sim&
-\frac{3}{4} [1 - 6 \Abar(G,W)][t + A(t)]\,\lnbar(\delta) ,
\\
K_{VVSV\Fbar\Fbar}(\delta, \delta, G, W, t, t) &\sim&
-\frac{3}{4t} [1 - 6 \Abar(G,W)][t + A(t)]\,\lnbar(\delta) .
\eeq
The $\lnbar(\delta)$ contributions in each of
eqs.~(\ref{eq:lndeltatWA}) and (\ref{eq:lndeltatGWA}) again are seen to cancel.

All other possible $\lnbar(\delta)$ contributions are found to vanish at the level
of the functions given in {\tt SMV3functions.anc}, except for the following contributions
involving doubled photon propagators, $W$ bosons, and charged Goldstone bosons:
\beq
&& \frac{e^4}{4} \Bigl [
K_{{\rm gauge}}(\delta,\delta,W,W,W,W) 
+ 4 W K_{{\rm gauge},S}(\delta,\delta,W,W,W,G)
\nonumber \\
&& 
+ 4 W^2 K_{VVSVVS}(\delta,\delta,G,W,W,G)
\Bigr ] .
\label{eq:V3lnbardelpart}
\eeq
The relevant infrared behaviors can again be extracted from {\tt SMV3functions.anc}: 
\beq
K_{{\rm gauge}}(\delta,\delta,W,W,W,W) &\sim& \frac{3}{16} [W + 6 A(W)]^2\, \lnbar(\delta)
\\
K_{{\rm gauge},S}(\delta,\delta,W,W,W,G) &\sim& 
-\frac{3}{32} [W + 6 A(W)] [1 - 6 \Abar(W,G)] \, \lnbar(\delta)
\\
K_{VVSVVS}(\delta,\delta,G,W,W,G) &\sim&
\frac{3}{64} [1 - 6 \Abar(W,G)]^2 \, \lnbar(\delta)
\eeq
It follows that the infrared divergences from doubled photon propagators do {\em not}
cancel. The final result for the infrared divergence, which comes entirely
from eq.~(\ref{eq:V3lnbardelpart}), can be simplified to:
\beq 
V^{(3)} &\sim& 
\frac{27 e^4}{16} 
\left (\frac{W G \ln(W/G)}{W-G}\right )^2 
\lnbar(\delta)
.
\label{eq:V3lnbardelta}
\eeq
While it might at first seem surprising that there is an uncanceled QED 
infrared divergence in the three-loop effective potential, 
it is important to remember that the effective potential itself is not a physical observable. (Recall that it is not even 
gauge invariant.) What is important is that this infrared divergence does not infect 
physical observables and closely related quantities. The key property that 
guarantees this is that eq.~(\ref{eq:V3lnbardelta}) is of second order in $G$.
As we will see in the next section, after the necessary resummation of Goldstone boson 
contributions, terms of higher-than-linear order in $G$ do not affect 
the minimization condition of the effective potential, nor contribute to the value of the effective potential at its minimum, and so can simply be dropped.

As an aside, 
one could also eliminate the infrared divergence in eq.~(\ref{eq:V3lnbardelta}) by 
resumming photon self-energies. Note that eq.~(\ref{eq:V3lnbardelta})
comes from the 3-loop representatives of the family
of $\ell$-loop Feynman diagrams that 
involve a ring of $\ell-1$ photon propagators that carry the same momentum
and connect $\ell-1$ one-loop subdiagrams with either $W^+,W^-$ or $W^\pm,G^\mp$
internal lines. 
These diagrams scale like $1/\delta^{\ell - 3}$ for $\ell > 3$.
Resumming these diagrams to all orders in $\ell$ would yield a contribution 
to $V_{\rm eff}$ that 
cancels the three-loop infrared divergence in eq.~(\ref{eq:V3lnbardelta}) 
in the limit $\delta \rightarrow 0$:
\beq
\Delta V_{\rm eff} &\sim& \frac{3}{16 \pi^2} f_V(\Delta_\gamma/16\pi^2)
- \frac{1}{(16\pi^2)^3} \frac{3}{4} \Delta_\gamma^2 [\lnbar(\delta) + 2/3]  + \ldots 
\eeq
where the ellipses represent contributions from four loops and beyond, and
\beq
\Delta_\gamma = 3 e^2 W G \ln(W/G)/2(W-G).
\eeq
However, this resummation of photon self-energies is actually an unnecessary complication. 
The key fact is that with or without the photon ring resummation 
the contribution is second order in $G$, and so has no effect on physical quantities
and can be dropped after Goldstone boson resummation. Therefore, I prefer not to
resum the photon self-energies, for simplicity.
In the next subsection, I will discuss the expansion 
and resummation of the Goldstone
boson contributions, and explicitly derive the resulting Higgs VEV minimization condition
and find that it has no infrared divergences [or spurious imaginary parts
associated with $\lnbar(G)$ when $G$ is negative] of any kind.

As a check of the Standard Model result, consider the 
renormalization group scale invariance conditions, 
which take the form
of eq.~(\ref{eq:QdQVexpanded}) at each loop order $\ell = 1,2,3$, 
where $X$  is summed over $\Lambda, m^2, \lambda, g_3, g, g', y_t, \phi$, 
with $\beta_\phi^{(\ell)} = -\gamma^{(\ell)}\phi$ where $\gamma$ is the scalar field
anomalous dimension. The necessary $Q$ derivatives of the three-loop integral functions
are given in the ancillary electronic file {\tt QdQ.anc}.
The necessary three-loop
beta functions and anomalous dimension 
have been found in 
refs.~\cite{Machacek:1983tz}-\cite{Bednyakov:2013eba},\footnote{Extensions to 
QCD 4-loop and 5-loop order will not be needed here, but can be 
found in refs.~\cite{vanRitbergen:1997va}-\cite{Chetyrkin:2016ruf}
and \cite{Martin:2015eia}.}  
except for the
field-independent vacuum energy $\Lambda$. From the $\phi$-independent parts of
eq.~(\ref{eq:QdQVexpanded}), I find that:
\beq
\beta_\Lambda^{(1)} &=& 2 (m^2)^2,
\\
\beta_\Lambda^{(2)} &=& [12 g^2 + 4 g^{\prime 2} - 12 y_t^2] (m^2)^2,
\\
\beta_\Lambda^{(3)} &=& \Bigl [ 
(192 \zetathree -204) y_t^2 g_3^2
+ 153 y_t^4/4 
+ (189/8 - 108 \zetathree) y_t^2 g^2
- (73/8 + 20 \zetathree) y_t^2 g^{\prime 2}
\nonumber \\ &&
+ (4215/32 - 51 n_G/2 - 18 \zetathree) g^4
+ (36 \zetathree -441/16) g^2 g^{\prime 2}
\nonumber \\ &&
+ (6 \zetathree -233/32 - 85 n_G/6 ) g^{\prime 4}
+ 18 \lambda^2
\Bigr ] (m^2)^2 .
\eeq
With this included, I have checked, using the $Q$ derivatives found in {\tt QdQ.anc},
that eq.~(\ref{eq:QdQVexpanded}) is indeed
satisfied by the result for $V^{(3)}$ given in 
{\tt SMV3.anc} and {\tt SMV3functions.anc}. 
[Note that the coefficient of $\lnbar(\delta)$ in eq.~(\ref{eq:V3lnbardelta})
is independent of $Q$, so that while eq.~(\ref{eq:V3lnbardelta}) 
does contribute non-trivially to $Q \partial V^{(3)}/\partial Q$,
it does so without producing a $\lnbar(\delta)$ part.]
This constitutes an important consistency check.

\subsection{Goldstone boson resummation of the Standard Model effective 
potential\label{subsec:Goldstoneresummation}}

The three-loop Standard Model effective potential given in the previous subsection
suffers from two related problems associated with the Goldstone boson contributions. 
First, the squared mass $G$ can easily be negative at the minimum of the 
(real part of the) effective potential, depending on the 
choice of renormalization scale $Q$ at which it is evaluated. Due 
to the presence of $\lnbar(G)$, the usual effective potential then 
has an imaginary part even at one-loop order. This imaginary part is spurious because 
it does not correspond to a genuine physical instability. Second,
if one chooses a reasonable renormalization 
scale such that $G \rightarrow 0$, then there are
$\lnbar(G)$ singularities in the three-loop effective potential and in the derivative
of the two-loop effective potential, and $1/G^{L-3}$
singularities in the $L$-loop effective potential and the derivatives of the
$(L-1)$-loop effective potential for $L > 3$.  
This was noted in the context of the Standard Model at leading order in the
top-quark Yukawa coupling in ref.~\cite{Martin:2013gka} 
(see also \cite{Ford:1992mv,Einhorn:2007rv}), 
where it was somewhat 
melodramatically referred to as the ``Goldstone boson catastrophe". 
In practice, one can usually simply ignore the problem while maintaining good numerical
accuracy, by choosing a renormalization scale such that $|G|$ is not too tiny, 
and dropping the imaginary part if $G$ is negative.
However, this is clearly sub-optimal, and a principled solution was given 
in refs.~\cite{Martin:2014bca,Elias-Miro:2014pca}, where
the problem was shown to be resolved by resumming the leading Goldstone boson 
contributions to all orders, treating $G$ as small compared to the other squared mass 
parameters of the theory. 

The basic idea is very simple: the effects of Goldstone
bosons with propagators with squared masses $G$ are re-expanded about a different squared 
mass $G+\Delta$, which vanishes at the minimum of the full effective potential; 
this is the pole squared mass of the Goldstone boson in Landau gauge, 
and therefore a good expansion point.
This resolution by resummation
has the added benefit that it actually makes it simpler in practice
to implement the minimization condition that relates the Higgs VEV to the Lagrangian
parameters. It can, and should, also be applied to other calculations such as the 
pole squared masses of the physical particles.
For important further developments and related perspectives, see 
refs.~\cite{Pilaftsis:2015cka}-\cite{Braathen:2017izn}.

To apply the Goldstone boson 
resummation procedure to the full 
three-loop\footnote{The extension of this whole procedure
to any given higher loop order should be clear from the following.} 
Standard Model effective potential,
consider the ordinary effective potential in the form:
\beq
V_{\rm eff} &=& V^{(0)} 
+ \frac{1}{16 \pi^2} V^{(1)}(G) 
+ \frac{1}{(16 \pi^2)^2} V^{(2)}(G)
+ \frac{1}{(16 \pi^2)^3} V^{(3)}(G).
\eeq
Here the dependence of each term 
on the Goldstone boson squared mass $G$ has now been indicated
explicitly, with the dependences on the other independent parameters 
$(g_3, g, g', y_t, \lambda, \phi^2, Q)$ left 
implicit.\footnote{Note that $H = 2 \lambda \phi^2 + G$, so that it is not an independent
parameter, and in the following
$\partial H/\partial G = 1$. The other squared mass parameters $t,W,Z$ are independent of
$G$.} 
Now one can resum the contributions to all loop orders
from diagrams that consist of single rings of Goldstone boson propagators punctuated by
one-particle-irreducible subdiagrams that feature larger masses, by writing
\beq
V^{(1)}(G) \rightarrow V^{(1)}(G + \Delta) 
- \Delta \frac{\partial}{\partial G} V^{(1)}(G)
- \frac{1}{2} \Delta^2 \frac{\partial^2}{\partial G^2} V^{(1)}(G)
\eeq
where the quantity 
\beq
\Delta = \frac{1}{16 \pi^2} \Delta_1 + \frac{1}{(16 \pi^2)^2} \Delta_2 +
\frac{1}{(16 \pi^2)^3} \Delta_3 + \ldots
\eeq
will be given below, and is defined\footnote{A warning about a notational switch: 
the $\Delta_\ell$ in the present paper 
are equal to what I called $\widehat \Delta_\ell$ in ref.~\cite{Martin:2014bca}.
The following discussion could be equally well reformulated in terms of the quantities
called $\Delta_\ell$ in ref.~\cite{Martin:2014bca}, 
with results that are consistent up to four-loop order contributions. 
However,
that alternative formulation is complicated slightly by the fact that the $\Delta_\ell$
in the notation of ref.~\cite{Martin:2014bca}
depend on $G$, through their dependences on $H$, so I omit it for simplicity.} 
by the properties that 
$G + \Delta$ vanishes at the minimum of the 
full effective potential, and
each $\Delta_\ell$ does not depend on $G$. 
Now we can write, through three-loop order:
\beq
V_{\rm eff} &\rightarrow& V^{(0)} 
+ \frac{1}{16 \pi^2} V^{(1)}(G + \Delta) 
+ \frac{1}{(16 \pi^2)^2} \widehat V^{(2)}(G)
\nonumber \\ &&
+ \frac{1}{(16 \pi^2)^3} 
\left [V^{(3)}(G) - \Delta_2 \frac{\partial}{\partial G} V^{(1)}(G)
- \frac{1}{2} (\Delta_1)^2 \frac{\partial^2}{\partial G^2} V^{(1)}(G) \right ]
,
\eeq
where we have defined
\beq
\widehat V^{(2)}(G) &\equiv& V^{(2)}(G) - \Delta_1 \frac{\partial}{\partial G} V^{(1)}(G),
\label{eq:defV2Ghat}
\eeq
Now one can continue the resummation procedure by making the replacement:
\beq
\widehat V^{(2)}(G) &\rightarrow& 
\widehat V^{(2)}(G+\Delta) - \Delta_1 \frac{\partial}{\partial G} \widehat V^{(2)}(G)
\nonumber \\ 
&=& 
\widehat V^{(2)}(G+\Delta) - \Delta_1 \frac{\partial}{\partial G} V^{(2)}(G)
+ (\Delta_1)^2 \frac{\partial^2}{\partial G^2} V^{(1)}(G)
,
\phantom{xxx}
\eeq
with the result
\beq
V_{\rm eff} &\rightarrow& V^{(0)} 
+ \frac{1}{16 \pi^2} V^{(1)}(G + \Delta) 
+ \frac{1}{(16 \pi^2)^2} \widehat V^{(2)}(G + \Delta)
+ \frac{1}{(16 \pi^2)^3} \widehat V^{(3)}(G)
,
\eeq
where
\beq
\widehat V^{(3)}(G)
&\equiv& V^{(3)}(G) 
- \Delta_1  \frac{\partial}{\partial G} V^{(2)}(G)
- \Delta_2 
\frac{\partial}{\partial G} V^{(1)}(G) 
+ \frac{1}{2} (\Delta_1)^2  \frac{\partial^2}{\partial G^2} V^{(1)}(G)
.
\phantom{xxx}
\label{eq:defV3Ghat}
\eeq
Finally, we can replace $G$ by $G+\Delta$ in the three-loop term, since the difference
is of four-loop order.
Thus, the resummed effective potential at three-loop order is:
\beq
V_{\rm eff}^{\rm resummed} = V^{(0)} 
+ \frac{1}{16 \pi^2} V^{(1)}(G + \Delta) 
+ \frac{1}{(16 \pi^2)^2} \widehat V^{(2)}(G + \Delta)
+ \frac{1}{(16 \pi^2)^3} \widehat V^{(3)}(G + \Delta)
,
\phantom{xxx}
\label{eq:SMV3resummedform}
\eeq
where the functions $\widehat V^{(2)}$ and 
$\widehat V^{(3)}$ are defined in terms of the usual
perturbatively calculated (non-resummed) quantities by eqs.~(\ref{eq:defV2Ghat}) and 
(\ref{eq:defV3Ghat}), respectively. 

In order to construct the functions
$\widehat V^{(2)}(G)$ and $\widehat V^{(3)}(G)$ from the 
results given in the previous subsection, 
one needs $\Delta_1$, $\Delta_2$, 
$\frac{\partial}{\partial G} V^{(1)}(G)$, 
$\frac{\partial^2}{\partial G^2} V^{(1)}(G)$, and
$\frac{\partial}{\partial G} V^{(2)}(G)$,
which are all straightforward
to obtain from the one-loop and two-loop order effective potentials.
The results for $\Delta_1$ and $\Delta_2$ have already been 
given in eqs.~(4.19) and (4.20)
of ref.~\cite{Martin:2014bca}, and are also provided in
an ancillary electronic file of the present paper called {\tt SMDeltas.anc}. 
For example, 
\beq
\Delta_1 &=& 3 \lambda A(h) -6 y_t^2 A(t) + \frac{3 g^2}{2} A(W) 
+ \frac{3(g^2 + g^{\prime 2})}{4} A(Z)  
\nonumber \\ &&
+ (3 g^4 + 2 g^2 g^{\prime 2} + g^{\prime 4}) \phi^2/8,\phantom{xxx}
\label{eq:Delta1}
\eeq
where
\beq
h \equiv H - G = 2 \lambda \phi^2.
\eeq
Also, one has the simple one-loop results:
\beq
\frac{\partial}{\partial G} V^{(1)}(G) &=& \frac{3}{2} A(G) + \frac{1}{2} A(H)
,
\label{eq:dV1dG}
\\
\frac{\partial^2}{\partial G^2} V^{(1)}(G) &=& 
\frac{3}{2} [1 + A(G)/G] + \frac{1}{2} [1 + A(H)/H]
.
\label{eq:d2V1dG2}
\eeq
The expression for $\partial V^{(2)}/\partial G$ is more complicated, and is
given in an ancillary electronic file {\tt SMdV2dG.anc}.

A crucial feature of $V_{\rm eff}^{\rm resummed}$ is that in the expansions of
$V^{(1)}(G+\Delta)$,
$\widehat V^{(2)}(G+\Delta)$, and $\widehat V^{(3)}(G+\Delta)$
for small $G+\Delta$, terms with $\lnbar(G+ \Delta)$
do not appear until quadratic order in $G+\Delta$. This can be seen by
performing the expansions for small $G$ for basis integral functions that have $G$
as an argument, using the tools in the ancillary electronic 
file {\tt expzero.anc}. The results 
can be written in the form
\beq
V^{(1)}(G) &=& V^{(1)}(0) + G V^{(1)\prime}(0) + {\cal O}(G^2),
\label{eq:V1Gexp}
\\
\widehat V^{(2)}(G) &=& \widehat V^{(2)}(0) + G \widehat V^{(2)\prime}(0) 
+ {\cal O}(G^2),\phantom{xx}
\label{eq:V2hatGexp}
\\
\widehat V^{(3)}(G) &=& \widehat V^{(3)}(0) + G \widehat V^{(3)\prime}(0) + {\cal O}(G^2)
,
\label{eq:V3hatGexp}
\eeq
where $V^{(1)}(0)$, $V^{(1)\prime}(0)$, $\widehat V^{(2)}(0)$, $\widehat V^{(2)\prime}(0)$, 
$\widehat V^{(3)}(0)$, and $\widehat V^{(3)\prime}(0)$ do not depend on $G$. 
In particular, the cancellations of the $G \lnbar(G)$ terms 
in $\widehat V^{(2)}(G)$, and the
$\lnbar(G)$, $G \lnbar(G)$, and $G \lnbar^2(G)$ terms in 
$\widehat V^{(3)}(G)$, provide an important check. Because of the absence of these
terms in eqs.~(\ref{eq:V2hatGexp}) and (\ref{eq:V3hatGexp}), 
the resummed effective potential $V_{\rm eff}^{\rm resummed}$
defined by eq.~(\ref{eq:SMV3resummedform}), 
and its first derivatives with respect to arbitrary parameters, are finite and real
at its minimum. 

Note also that the expansions to linear order given in 
eqs.~(\ref{eq:V1Gexp})-(\ref{eq:V3hatGexp}), applied to eq.~(\ref{eq:SMV3resummedform}),
are sufficient to produce the minimization condition for the
Higgs VEV valid through full three-loop order, because first derivatives of
terms of order $(G+\Delta)^2$ or higher will vanish there. Since the quadratic terms
have been dropped, the QED infrared divergence of eq.~(\ref{eq:V3lnbardelta})
does not appear.

The explicit one-loop and two-loop order results are
\beq
V^{(1)}(0) &=& f(h) - 12 f(t) + 6 f_V(W) + 3 f_V(Z),
\\
V^{(1)\prime}(0) &=& A(h)/2,
\eeq
and
\beq
\widehat V^{(2)}(0) &=& 
\frac{3}{4} \lambda f_{SS}(h,h) 
+ 3 \lambda^2 \phi^2 [f_{SSS}(h,h,h) 
+ f_{SSS}(0,0,h)]
+ \frac{3 y_t^2}{2} \Bigl [
f_{FFS}(t,t,h) 
\nonumber \\ &&
+ t f_{\Fbar\Fbar S}(t,t,h)
+ f_{FFS}(t,t,0) - t f_{\Fbar\Fbar S}(t,t,0)
+ 2 f_{FFS}(0,t,0)
\Bigr ]
\nonumber \\ &&
+ \frac{g^2 + \gptwo}{8} f_{VSS}(Z,0,h)
+ \frac{(g^2 - \gptwo)^2}{8 (g^2 + \gptwo)} f_{VSS}(Z,0,0)
\nonumber \\ &&
+ \frac{g^2}{4} \left [ f_{VSS}(W,0,h) + f_{VSS}(W,0,0) \right]
+ \frac{g^4 \phi^2}{8} f_{VVS}(W,W,h)
\nonumber \\ &&
+ \frac{(g^2 + \gptwo)^2 \phi^2}{16} f_{VVS}(Z,Z,h)
+\frac{e^2 \phi^2}{4}\left [\gptwo f_{VVS}(W,Z,0)
+ g^2 f_{VVS}(0,W,0)\right ]
\nonumber \\ &&
-\left [4 g_3^2 + 4 e^2/3
            \right ] t f_{\overline{F}\overline{F}V}(t,t,0)
+ g^2 \left [ 3 f_{FFV}(0,t,W) + (4 n_G - 3) f_{FFV}(0,0,W) \right ]/2
\nonumber \\ &&
+ \bigl [
(9 g^4 - 6 g^2 \gptwo + 17 \gpfour) f_{FFV}(t,t,Z)
+  8 \gptwo (3 g^2 - \gptwo) t f_{\overline{F}\overline{F}V}(t,t,Z)
\nonumber \\ &&
+ ((24 n_G -9) g^4 + 6 g^2 \gptwo + (40 n_G - 17) \gpfour) f_{FFV}(0,0,Z)
\bigr ]/24 (g^2 + \gptwo)
\nonumber \\ &&
+ \frac{e^2}{2} f_{\rm gauge}(W,W,0)
+ \frac{g^4}{2 (g^2 + \gptwo)} f_{\rm gauge}(W,W,Z)
- \Delta_1 A(h)/2,
\label{eq:V2SM0}
\\
\widehat V^{(2)\prime}(0) &=& 
3 y_t^2 (y_t^2/4\lambda - 1) I(h,t,t) 
+ 3 [g^2/4 - \lambda - g^4/32\lambda + g^4/16 (g^2 - 2\lambda)] I(h,W,W) 
\nonumber \\ &&
+ 3 [(g^2 + g^{\prime 2})/8 - \lambda/2 - (g^2 + g^{\prime 2})^2/64 \lambda
+ (g^2 + g^{\prime 2})^2/32 (g^2 + g^{\prime 2} - 2 \lambda)] I(h,Z,Z)
\nonumber \\ &&
- 3 \lambda I(h,h,h)/2
+ (6 \lambda + 3 g^2/4) I(0,h,W) 
+ [3 \lambda + 3 (g^2 + g^{\prime 2})/8] I(0,h,Z)
\nonumber \\ &&
+ [3 (2 g^2 + g^{\prime 2})^3/4 (g^2 + g^{\prime 2})^2] I(0,W,Z)
+ \Bigl \{ 
3 \bigl [-(g^2 + g^{\prime 2})/16 \lambda
\nonumber \\ &&
- (g^2 + g^{\prime 2})/8(g^2 + g^{\prime 2} - 2\lambda)
+ (g^{\prime 4} -g^4 + 6 g^2 g^{\prime 2})/4(g^2 + g^{\prime 2})^2
\bigr ]  A(Z)^2
\nonumber \\ &&
-3 \bigl [g^2/8\lambda + g^2/4(g^2 - 2 \lambda) 
+ (g^4 + 6 g^2 g^{\prime 2} + 4 g^{\prime 4})/2(g^2 + g^{\prime 2})^2 
\bigr ] A(W)^2
\nonumber \\ &&
+ [3 (8 g^4 + 8 g^2 g^{\prime 2} + g^{\prime 4})/(g^2 + g^{\prime 2})^2] 
A(W) A(Z)
- (3 y_t^2/2\lambda) A(h) A(t)
\nonumber \\ &&
+ [3 (g^2 + g^{\prime 2})^2/16 \lambda (g^2 + g^{\prime 2} - 2 \lambda)] 
A(h) A(Z)
+ [3 g^4/8 \lambda (g^2 - 2 \lambda)] A(h) A(W)
\nonumber \\ && 
+ (9 + 3 y_t^2/2\lambda) A(t)^2 + 3 A(h)^2/4
\Bigr \}/\phi^2
- (3 y_t^4/2\lambda) A(t)
\nonumber \\ &&
+ 3 [(g^2 + g^{\prime 2})^2 (g^2 + g^{\prime 2}- 4\lambda)/
  32\lambda (g^2 + g^{\prime 2}- 2\lambda)
\nonumber \\ &&
-g^2 (7 g^2 + 10 g^{\prime 2})/4 (g^2 + g^{\prime 2})] A(Z)
\nonumber \\ &&
+ 3 g^2 [g^2/16\lambda - g^2/8 (g^2 - 2 \lambda) 
- (6 g^4 + 6 g^2 g^{\prime 2} + g^{\prime 4})/4(g^2 + g^{\prime 2})^2] A(W)
\nonumber \\ &&
+ [-3y_t^4/4\lambda  + 3 y_t^2/2 - 3\lambda -3g^2/8 - g^{\prime 2}/8
+ (21 g^4 + 14 g^2 g^{\prime 2} + 7 g^{\prime 4})/64\lambda
\nonumber \\ &&
- 3 (g^2 + g^{\prime 2})^2/32 (g^2 + g^{\prime 2} - 2 \lambda)
- 3 g^4/16 (g^2 - 2 \lambda)
] A(h)
\nonumber \\ &&
+ [
3 y_t^4 + 6 \lambda^2 - 9 g^4/8 + 3 g^2 g^{\prime 2}/8 + 
3 g^6/8 (g^2 + g^{\prime 2})
 + 3 g^8/16 (g^2 + g^{\prime 2})^2
] \phi^2
\zetatwo
\nonumber \\ &&
+ 3[y_t^6/4\lambda  + y_t^4 - \lambda y_t^2 + \lambda (3 g^2 + g^{\prime 2})/6
- (3 g^6 + 3 g^4 g^{\prime 2} + 3 g^2 g^{\prime 4} + g^{\prime 6})/128\lambda
\nonumber \\ &&
+ 3 g^6/32 (g^2 - 2 \lambda) 
+ 3 (g^2 + g^{\prime 2})^3/64 (g^2 + g^{\prime 2} - 2 \lambda)
+ (81 g^8 + 158 g^6 g^{\prime 2} 
\nonumber \\ &&
+ 110 g^4 g^{\prime 4} + 28 g^2 g^{\prime 6} + 
g^{\prime 8})/96 (g^2 + g^{\prime 2})^2] \phi^2
\eeq
These are included, along with the much more complicated results for
$\widehat V^{(3)}(0)$ and $\widehat V^{(3)\prime}(0)$,
in an ancillary electronic file {\tt SMVresummedGexp.anc}.
The results are given in terms
of basis loop integral functions, with squared mass arguments $h,t,W,Z,0$ and
with coefficients built out of $g_3, g, g', y_t, \lambda$, and $\phi^2$.

\subsection{The Standard Model Higgs VEV at three-loop order\label{subsec:SMV3min}}

In this subsection, I discuss the application of the Standard Model 
three-loop effective potential to obtain the minimization condition for the Higgs
VEV $v = \phi_{\rm min}$, 
given the Higgs squared mass parameter $m^2$, or vice versa. This condition is
\beq
\frac{1}{\phi} \frac{\partial}{\partial \phi} 
V^{\rm resummed}_{\rm eff}\biggl |_{\phi = v} &=& 0,
\eeq
which can be written  as
\beq
G = m^2 + \lambda v^2 = 
-\sum_{\ell = 1}^\infty \frac{1}{(16\pi^2)^\ell} \Delta_\ell
.
\label{eq:GminDelta}
\eeq
This result can also be expressed as the relation between the \MSbar tree-level VEV
\beq
v_{\rm tree} \equiv \sqrt{-m^2/\lambda}
\eeq
and the VEV $v$ defined as the minimum of the full effective potential. One has:
\beq
v_{\rm tree}^2 = v^2 + \frac{1}{\lambda} \sum_{\ell = 1}^{\infty} \frac{1}{(16\pi^2)^\ell} \Delta_\ell 
.
\eeq
Using the expansions of eq.~(\ref{eq:V1Gexp})-(\ref{eq:V3hatGexp}) 
in eq.~(\ref{eq:SMV3resummedform})
and the fact that, by definition, $G + \Delta$ vanishes at the minimum, we have
through three-loop order:
\beq
\Delta_1 &=& \frac{1}{\phi} \frac{\partial}{\partial \phi} V^{(1)}(0) 
+ V^{(1)\prime}(0) \frac{1}{\phi} \frac{\partial G}{\partial \phi}
,
\\
\Delta_2 &=& \frac{1}{\phi} \frac{\partial}{\partial \phi} \widehat V^{(2)}(0) + 
\widehat V^{(2)\prime}(0) \frac{1}{\phi}\frac{\partial G}{\partial \phi} 
+ V^{(1)\prime}(0) \frac{1}{\phi}\frac{\partial \Delta_1}{\partial \phi}
,
\\
\Delta_3 &=& \frac{1}{\phi} \frac{\partial}{\partial \phi} \widehat V^{(3)}(0) 
+ \widehat V^{(3)\prime}(0) \frac{1}{\phi}\frac{\partial G}{\partial \phi}
+ \widehat V^{(2)\prime}(0) \frac{1}{\phi}\frac{\partial \Delta_1}{\partial \phi}
+ V^{(1)\prime}(0) \frac{1}{\phi}\frac{\partial \Delta_2}{\partial \phi}
,
\eeq
with $\phi = v$. Now one can use:
\beq
\frac{1}{\phi}\frac{\partial G}{\partial \phi} &=& 2 \lambda,
\\
\frac{1}{\phi}\frac{\partial \Delta_1}{\partial \phi} &=& 
[6 \lambda A(h) -12 y_t^2 A(t) + 3 g^2 A(W) + 3 (g^2 + g^{\prime 2}) A(Z)/2]/\phi^2
\nonumber \\ &&
+ 12 \lambda^2 - 6 y_t^4 + 15 g^4/8 + 5 g^2 g^{\prime 2}/4 + 5 g^{\prime 4}/8
\eeq
together with the derivatives of the two-loop and three-loop basis functions
as given in ref.~\cite{Martin:2016bgz}, 
to iteratively evaluate $\Delta_1$, $\Delta_2$, and $\Delta_3$. As mentioned above,
the first two were already given above in eqs.~(4.19) and (4.20) 
of ref.~\cite{Martin:2014bca}, and
all three are given in the ancillary electronic file {\tt SMDeltas.anc} distributed
with the present paper. These results are given in terms of basis functions with arguments
$h,t,W,Z,0$ and \MSbar renormalization scale $Q$. The complete lists of the specific 
one-loop, two-loop, and three-loop
basis functions needed are:
\beq
{\cal I}^{(1)} &=& \{A(h), A(t), A(W), A(Z)\},
\\
{\cal I}^{(2)} &=& \{
\zeta_2 
,
I(0, h, W), I(0, h, Z), I(0, t, W), I(0, W, Z), I(h, h, h), 
\nonumber \\ && 
I(h, t, t), I(h, W, W), I(h, Z, Z), I(t, t, Z), I(W, W, Z)
\},
\\
{\cal I}^{(3)} &=& \{
\zeta_3,
F(h, 0, 0, t), F(h, 0, 0, W), F(h, 0, 0, Z), 
F(h, 0, h, W), F(h, 0, h, Z), F(h, 0, t, t), 
\nonumber \\ && 
F(h, 0, t, W), F(h, 0, W, W), F(h, 0, W, Z), 
F(h, 0, Z, Z), F(h, h, t, t), F(h, h, W, W), 
\nonumber \\ && 
F(h, h, Z, Z), F(h, t, t, Z), F(h, W, W, Z), 
F(t, 0, 0, W), F(t, 0, 0, Z), F(t, 0, h, W), 
\nonumber \\ && 
F(t, 0, t, W), F(t, 0, W, Z), F(t, h, t, Z), 
F(t, t, W, W), F(t, t, Z, Z), F(W, 0, 0, Z), 
\nonumber \\ && 
F(W, 0, h, h), F(W, 0, h, t), F(W, 0, h, Z), 
F(W, 0, t, t), F(W, 0, t, Z), F(W, 0, W, W), 
\nonumber \\ && 
F(W, 0, Z, Z), F(W, h, W, Z), F(Z, 0, h, h), 
F(Z, 0, h, W), F(Z, 0, t, t), F(Z, 0, t, W), 
\nonumber \\ && 
F(Z, 0, W, W), F(Z, 0, W, Z), F(Z, 0, Z, Z), 
F(Z, h, t, t), F(Z, h, W, W), \Fbar(0, 0, h, t), 
\nonumber \\ && 
\Fbar(0, 0, h, W), \Fbar(0, 0, h, Z), \Fbar(0, 0, t, W), 
\Fbar(0, 0, t, Z), \Fbar(0, 0, W, Z), \Fbar(0, h, t, t), 
\nonumber \\ && 
\Fbar(0, h, W, W), \Fbar(0, t, t, Z), \Fbar(0, W, W, Z), 
 G(0, 0, 0, h, t), G(0, 0, 0, t, Z), 
\nonumber \\ && 
G(0, t, t, W, W), G(h, 0, 0, 0, W), G(h, 0, 0, 0, Z), 
G(h, 0, 0, h, h), G(h, 0, 0, t, t), 
\nonumber \\ && 
G(h, 0, 0, W, W), G(h, 0, 0, Z, Z), G(h, 0, W, 0, Z), 
G(h, 0, W, h, h), G(h, 0, W, t, t), 
\nonumber \\ && 
G(h, 0, W, W, W), G(h, 0, W, Z, Z), G(h, 0, Z, h, h), 
G(h, 0, Z, t, t), G(h, 0, Z, W, W), 
\nonumber \\ && 
G(h, 0, Z, Z, Z), G(h, h, h, h, h), G(h, h, h, t, t), 
G(h, h, h, W, W), G(h, h, h, Z, Z), 
\nonumber \\ && 
G(h, t, t, W, W), G(h, t, t, Z, Z), G(h, W, W, Z, Z), 
G(t, 0, 0, 0, W), G(t, 0, 0, h, t), 
\nonumber \\ && 
G(t, 0, 0, t, Z), G(t, 0, W, h, t), G(t, 0, W, t, Z), 
G(t, h, t, t, Z), G(W, 0, 0, 0, Z), 
\nonumber \\ && 
G(W, 0, 0, h, W), G(W, 0, 0, W, Z), G(W, 0, h, 0, t), 
G(W, 0, h, 0, Z), G(W, 0, h, h, W), 
\nonumber \\ && 
G(W, 0, h, W, Z), G(W, 0, t, 0, Z), G(W, 0, t, h, W), 
G(W, 0, t, W, Z), G(W, 0, Z, h, W), 
\nonumber \\ && 
G(W, 0, Z, W, Z), G(W, h, W, W, Z), G(Z, 0, 0, h, Z), 
G(Z, 0, 0, t, t), G(Z, 0, 0, W, W), 
\nonumber \\ && 
G(Z, 0, h, 0, W), G(Z, 0, h, h, Z), G(Z, 0, h, t, t), 
G(Z, 0, h, W, W), G(Z, 0, W, h, Z), 
\nonumber \\ && 
G(Z, 0, W, t, t), G(Z, 0, W, W, W), G(Z, h, Z, t, t), 
G(Z, h, Z, W, W), G(Z, t, t, W, W),
\nonumber \\ && 
 H(0, 0, 0, 0, 0, h), H(0, 0, 0, 0, 0, t), 
 H(0, 0, 0, 0, 0, W), H(0, 0, 0, 0, 0, Z), 
\nonumber \\ && 
H(0, 0, 0, 0, h, W), H(0, 0, 0, 0, t, t), 
H(0, 0, 0, 0, t, W), H(0, 0, 0, 0, W, W), 
\nonumber \\ && 
H(0, 0, 0, 0, W, Z), H(0, 0, 0, h, t, t), 
H(0, 0, 0, h, W, W), H(0, 0, 0, h, Z, Z), 
\nonumber \\ && 
H(0, 0, 0, t, t, Z), H(0, 0, 0, W, W, Z), 
H(0, 0, h, 0, W, W), H(0, 0, h, h, W, W), 
\nonumber \\ && 
H(0, 0, h, h, Z, Z), H(0, 0, h, W, 0, 0), 
H(0, 0, h, W, 0, Z), H(0, 0, h, W, W, Z), 
\nonumber \\ && 
H(0, 0, h, Z, 0, 0), H(0, 0, h, Z, 0, W), 
H(0, 0, h, Z, W, W), H(0, 0, t, 0, t, W), 
\nonumber \\ && 
H(0, 0, t, 0, W, W), H(0, 0, t, h, t, t), 
H(0, 0, t, W, 0, t), H(0, 0, t, Z, 0, 0), 
\nonumber \\ && 
H(0, 0, t, Z, 0, W), H(0, 0, t, Z, W, W), 
H(0, 0, W, 0, W, Z), H(0, 0, W, h, h, h), 
\nonumber \\ && 
H(0, 0, W, h, Z, Z), H(0, 0, W, t, h, t), 
H(0, 0, W, t, t, Z), H(0, 0, W, W, 0, 0), 
\nonumber \\ && 
H(0, 0, W, W, 0, h), H(0, 0, W, W, 0, W), 
H(0, 0, W, W, 0, Z), H(0, 0, W, W, h, W), 
\nonumber \\ && 
H(0, 0, W, W, W, Z), H(0, 0, W, Z, 0, 0), 
H(0, 0, W, Z, 0, h), H(0, 0, W, Z, h, Z), 
\nonumber \\ && 
H(0, 0, W, Z, t, t), H(0, 0, Z, h, h, h), 
H(0, 0, Z, h, W, W), H(0, 0, Z, W, h, W), 
\nonumber \\ && 
H(0, 0, Z, Z, 0, 0), H(0, 0, Z, Z, 0, W), 
H(0, 0, Z, Z, W, W), H(0, h, h, W, h, W), 
\nonumber \\ && 
H(0, h, h, Z, h, Z), H(0, h, t, Z, t, t), 
H(0, h, W, W, W, h), H(0, h, W, W, W, Z), 
\nonumber \\ && 
H(0, h, Z, W, Z, W), H(0, h, Z, Z, Z, h), 
H(0, t, t, t, 0, t), H(0, t, t, t, h, t), 
\nonumber \\ && 
H(0, t, t, t, Z, t), H(0, t, t, W, 0, W), 
H(0, t, t, W, h, W), H(0, t, t, W, Z, W), 
\nonumber \\ && 
H(0, W, W, W, 0, W), H(0, W, W, W, h, W), 
H(0, W, W, W, Z, W), H(0, W, W, Z, h, Z), 
\nonumber \\ && 
H(0, W, Z, Z, W, W), H(h, h, h, h, h, h), 
H(h, h, t, h, t, t), H(h, h, W, h, W, W), 
\nonumber \\ && 
H(h, h, Z, h, Z, Z), H(h, t, t, t, h, t), 
H(h, t, t, t, Z, t), H(h, t, Z, t, t, Z), 
\nonumber \\ && 
H(h, W, W, W, h, W), H(h, W, W, W, Z, W), 
H(h, W, Z, W, W, Z), 
\nonumber \\ && 
H(h, Z, Z, Z, h, Z), 
H(t, t, Z, Z, t, t), H(W, W, Z, Z, W, W)
\} .
\eeq
(Recall that this choice of basis functions is not unique, because of the identities
mentioned at the end of section \ref{sec:setup}.)
The form of the result is then
\beq
\Delta_3 &=& 
\sum_i c_i^{(3)} {\cal I}_i^{(3)} +
\sum_{i,j} c_{i,j}^{(2,1)} {\cal I}_i^{(2)} {\cal I}_j^{(1)} +
\sum_{i} c_{i}^{(2)} {\cal I}_i^{(2)}  +
\sum_{i,j,k} c_{i,j,k}^{(1,1,1)} 
   {\cal I}_i^{(1)} {\cal I}_j^{(1)} {\cal I}_k^{(1)}
\nonumber \\ &&
+
\sum_{i,j} c_{i,j}^{(1,1)} 
   {\cal I}_i^{(1)} {\cal I}_j^{(1)} +
\sum_{i} c_{i}^{(1)} 
   {\cal I}_i^{(1)} +
c^{(0)},
\eeq
with coefficients that are built out of $g_3$, $g$, $g'$, $y_t$, $\lambda$, and
$\phi^2$, and are given in {\tt SMDeltas.anc}.
The result for $\Delta_3$ extends the partial result (in the approximation that
$g_3^2, y_t^2 \gg g^2,g^{\prime 2},\lambda$) given in eq.~(4.21) of
ref.~\cite{Martin:2014bca}.
The expression for $\Delta_4$ is known at leading order in QCD only,
and was given in eq.~(5.5) [see also eqs.~(4.39) and (4.40)] of 
ref.~\cite{Martin:2015eia}.
 
As a check, one can demand that eq.~(\ref{eq:GminDelta}) 
satisfies renormalization group scale invariance. 
This condition takes the form, for each loop order $\ell$:
\beq
Q \frac{\partial}{\partial Q} \Delta_\ell 
&=&
-\phi^2 \beta^{(\ell)}_\lambda
- 2 \lambda \phi \beta^{(\ell)}_\phi
+ \lambda \phi^2 (\beta_{m^2}^{(\ell)}/m^2) +
\sum_{n=1}^{\ell-1}
\biggl[(\beta^{(n)}_{m^2}/m^2) - \sum_X \beta^{(n)}_X \frac{\partial}{\partial X} \biggr]
\Delta_{\ell-n}
\nonumber \\ &&
\label{eq:checkRGDeltas}
\eeq
where $\phi=v$ at the minimum of the potential, and $X$ is summed over $g_3$, $g$, $g'$, $y_t$, $\lambda$, and $\phi$. I have verified eq.~(\ref{eq:checkRGDeltas}) for each of
$\ell=1,2,3$, using the results above.

\section{Outlook\label{sec:Outlook}}
\setcounter{equation}{0}
\setcounter{figure}{0}
\setcounter{table}{0}
\setcounter{footnote}{1}

In this paper I have provided the results for the effective potential 
at full three-loop order for a general renormalizable theory,
in the \MSbar scheme and using Landau gauge fixing. 
The results for the Standard Model provided in section \ref{sec:SM} allow the most
accurate theoretical determination possible at this time for 
the relationship between the Higgs VEV 
and the Lagrangian parameters, including the negative Higgs squared mass parameter $m^2$.
In practice, this can be used to eliminate $m^2$ and $G$ in favor of $v$ 
(and $H = 3 \lambda v^2 + m^2$ in favor of $h=2\lambda v^2$), from other calculations in which they appear.
A study of the numerical impact of the three-loop contributions is not given here, 
but will appear in future work. This is also part of a larger program, as begun
in refs.~\cite{Martin:2014cxa,Martin:2015lxa,Martin:2015rea,Martin:2016xsp}, 
to obtain high-precision results for the pole masses of the Standard Model particles,
and other observables, in the tadpole-free pure \MSbar scheme.

In general, three-loop order contributions to the effective potential
can suffer from various kinds of infrared divergences that arise due to 
doubled propagators carrying the same momentum and small squared masses. 
The problematic contributions associated with Goldstone bosons are eliminated by
resummation. 
The infrared divergences associated with doubled gluon propagators cancel completely
after including all diagrams at three-loop order.
I also found an uncanceled
infrared divergence from doubled photon propagators in the three-loop Standard Model effective potential; this can be eliminated by resummation of photon self-energies,
but it is actually benign even without doing so, provided that one resums the 
Goldstone boson contributions. 

One might also worry about the case of doubled massless or light 
fermion propagators, for example in models of supersymmetry breaking such as 
the O'Raifeartaigh model \cite{ORaifeartaigh:1975nky} 
that feature massless goldstino fermions. 
However, the results above 
show explicitly that, as suggested by 
power counting arguments, there are no such infrared divergences 
from massless fermions (no ``goldstino catastrophe") at three-loop order.

The \MSbar renormalization scheme based on dimensional regularization 
does not respect supersymmetry when there are gauge fields present. 
Therefore, the results given here are not of 
direct applicability to softly broken supersymmetric gauge theories, such as realistic
supersymmetric extensions of the Standard Model. Further work will be needed
in order to obtain the three-loop effective potential in the \DRbarprime renormalization
scheme \cite{Jack:1994rk} based on regularization by
dimensional reduction \cite{Siegel:1979wq,Capper:1979ns,Jack:1997sr}, 
which is appropriate for such theories.

{\it Acknowledgments:} This work was supported in part by the National
Science Foundation grant number PHY-1417028. This work was performed in part at
the Aspen Center for Physics, which is supported by National Science Foundation 
grant PHY-1607611.


\end{document}